\documentclass[english]{article}
\pdfoutput=1
\usepackage[T1]{fontenc}
\usepackage[latin9]{inputenc}
\usepackage{amsmath}
\usepackage{amssymb}
\usepackage{graphicx}
\usepackage{esint}
\usepackage{color}
\usepackage{hyperref}

\newcommand{\eq}{\begin{equation}}
\newcommand{\eqx}{\end{equation}}

\newcommand{\sg}{\sigma}

\newcommand{\Dl}{\Delta}

\newcommand{\kap}{\kappa}
\newcommand{\dw}{\partial}
\newcommand{\dwb}{\bar{\partial}}

\newcommand{\f}[2]{\frac{#1}{#2}}

\makeatletter
\usepackage{a4wide}

\makeatother

\usepackage{babel}
\begin{document}

\title{Classical limit of diagonal form factors and HHL correlators }

\author{Zoltan Bajnok$^{a}$\thanks{e-mail: {\tt bajnok.zoltan@wigner.mta.hu}},\ \  
Romuald A. Janik$^{b}$\thanks{e-mail: {\tt romuald@th.if.uj.edu.pl}} \\ \\ 
\small 
${}^a$ MTA Lend\"ulet Holographic QFT Group\\\small
Wigner Research Centre\\\small
H-1525 Budapest 114, P.O.B. 49, Hungary\\\small
${}^b$ Institute of Physics\\\small
Jagiellonian University\\\small
ul. {\L}ojasiewicza 11, 30-348 Krak{\'o}w, Poland}

\date{}
\maketitle
\begin{abstract}
We propose an expression for the classical limit of diagonal form
factors in which we integrate the corresponding observable over the moduli
space of classical solutions. In infinite volume the integral has
to be regularized by proper subtractions and we present the one, which
corresponds to the classical limit of the connected diagonal form
factors. In finite volume the integral is finite and can be expressed
in terms of the classical infinite volume diagonal form factors and
subvolumes of the moduli space. We analyze carefully the periodicity
properties of the finite volume moduli space and found a classical
analogue of the Bethe-Yang equations. By applying the results to the
heavy-heavy-light three point functions we can express their strong
coupling limit in terms of the classical limit of the sine-Gordon
diagonal form factors.
\end{abstract}

\section{Introduction}

Integrable two dimensional quantum field theories are very special
as, in principle, they can be solved exactly by the bootstrap method.
This method consists of two parts: the S-matrix bootstrap calculates
the scattering matrix of the theory from global symmetries and from
such physical requirements as crossing symmetry and unitarity 
\cite{Mussardo:1992uc,Dorey:1996gd}.
The second step is the form factor bootstrap, which uses the already
calculated S-matrix to determine the matrix elements of local operators,
from which the correlation functions can be built up 
\cite{Smirnov:1992vz,Babujian:1998uw,Babujian:2003sc}.
This program has been pushed forward to many interesting theories including
the sine-Gordon and sinh-Gordon theories \cite{Smirnov:1986vy,Fring:1992pt}. 

In the last decade there has been increasing interest and relevant
progress in applying the bootstrap program for the AdS/CFT correspondence
\cite{Beisert:2010jr}. The S-matrix bootstrap was successfully implemented,
which eventually lead to the complete description of the spectral
problem. Recently the focus moved to the application of the form factor
bootstrap. An axiomatic approach for world-sheet form factors 
was developed in \cite{Klose:2013aza,Klose:2012ju}.
In \cite{Bajnok:2014sza} it was suggested that finite
volume diagonal form factors can be used to describe the Heavy-Heavy-Light
(HHL) 3-point functions. This proposal has been tested both at weak
\cite{Hollo:2015cda} and strong coupling and for special operators
\cite{Bajnok:2014sza}. Recently we also made a proposal, how the
form factor axioms can be modified to describe the string field theory
vertex, which corresponds to generic 3-point functions on the gauge
theory side \cite{Bajnok:2015hla}. This was complemented by the hexagon
approach \cite{Basso:2015zoa}, which were devised to calculate the
3-point functions directly by cutting the pant diagram into two hexagons.
These hexagons were exactly calculated and the method was checked
by comparing to weak coupling data \cite{Basso:2015eqa,Eden:2015ija}. 
Later it was shown that in the diagonal limit the results reproduce
the structure of the diagonal form factor proposal for HHL correlators
\cite{Jiang:2015bvm,Jiang:2016dsr}. The HHH three point functions
were also analyzed recently in \cite{Jiang:2016ulr}.

In testing the HHL proposal at
strong couplings a check for two particles was performed \cite{Bajnok:2014sza}.
We observed that in this limit the 3-point function was related to
the average of the light vertex operator over the moduli space of
classical solutions. As the strong coupling limit of the model is related to
the classical limit of the sine-Gordon theory it is natural to assume
that the classical limit of form factors are just the average of the
corresponding observable for the moduli space of classical solutions.
The aim of our paper is to investigate this correspondence.

Interestingly, there were not many investigations on the classical
limit of form factors. 
Goldstone and Jackiw \cite{Goldstone:1974gf,Rajaraman:1982is}
showed that the classical kink solution is the Fourier transform of
the form factor of the basic field between two moving kink states
in the semi-classical limit, when the kink momentum is very small
compared to its mass. Later Mussardo et al. extended the expression
into a transparently relativistically covariant form and used its
crossed version to determine the masses of boundstates \cite{Mussardo:2003ji}.
In the diagonal limit these analyses dictate that the classical limit
of the elementary fields' form factor between one-particle states
should be the spatial integral of the static kink solution. Our
work gives a meaning for this formula and generalizes the result for
generic operators and for multi-particle states. Let 
us also mention that semiclassical finite volume form factors were analyzed 
in \cite{Smirnov:1998kv} in the conformal case. Here, in contrast, 
we focus on massive scattering theories. In such a theory, namely in 
the sinh-Gordon theory, Lukyanov analyzed the semiclassical expansion 
of the finite temperature expectation values of exponential fields   \cite{Lukyanov:2000jp}. 
These results are valid for any volume but only for the vacuum expectation 
value. Here we deal with asymptotically large volumes and expectation 
values in multiparticle states. 

Our paper is organized as follows. In section~2 we give a brief heuristic
introduction to the paper. In section~3 we present our proposal for 
the classical computation of multiparticle diagonal form factors in
infinite volume. Then we move on in section~4 to describe the evaluation
of finite volume expectation values in the classical limit and establish
their link with the classical diagonal form factors of the previous
section. In section~5 we briefly comment on the link with Heavy-Heavy-Light
OPE coefficients and we close the paper with conclusions and two appendices.

\section{Guide to the paper}

Here we make a heuristic argument why the diagonal form factors should
be evaluated in the classical limit by averaging the operators for
the moduli space of classical solutions. Let us calculate the diagonal
form factor by evaluating the path integral:
\begin{equation}
\langle p_{1},...,p_{n}\vert\mathcal{O}(\phi(x,t))\vert p_{n},\dots,p_{1}\rangle
=\int_{\phi_{in}}^{\phi_{out}}d[\phi]\mathcal{O}(\phi(x,t))e^{\frac{i}{\hbar}S[\phi]}
\end{equation}
where the initial configuration, $\phi_{in}$, is related to a multiparticle
state with momenta $\{p_{i}\}$ prepared at $t\to-\infty$, while
the final configuration, $\phi_{out}$, is also a multiparticle state
with the same momenta $\{p_{i}\}$ fixed at $t\to\infty$. As the
particles' momenta are all different, for asymptotically large
times particles form well-separated non-interacting one-particle states.
There are many configurations with the prescribed momentum content,
$\{p_{i}\}$, all of which can be obtained by shifting the
trajectories of each of the asymptotic one-particle states, $\{x_{i}\}$.
These shifts do not effect the scattering matrix, but modify the path
integral and generate the moduli space of classical solutions. In
the classical limit ($\hbar\to0$) the path integral localizes exactly
to this moduli space
\begin{equation}
\langle p_{1},...,p_{n}\vert\mathcal{O}(\phi(x,t))\vert p_{n},\dots p_{1}\rangle
=\mathcal{N}\int_{\mathcal{M}}\prod dx_{i}\mathcal{O}(\phi_{n}(x,t,\{x_{i}\},\{p_{i}\}))
\label{e.secondtry}
\end{equation}
where $\phi_{n}$ is the classical $n$-particle solution with momenta
$\{p_{i}\}$ and shift parameters $\{x_{i}\}$, which form the classical
moduli space $\mathcal{M}$ and the normalization is proportional
to the action, which is constant on the moduli space: 
$\mathcal{N}\propto e^{\frac{i}{\hbar}S[\phi_{n}]}$.

The infinite volume moduli space is clearly noncompact and the relevant integral
is infinite as it stands.
This is in fact an
exact counterpart of the divergences of the infinite volume form factor in the diagonal limit which arise
due to disconnected pieces with smaller particle number (see section~3.1). The divergences in the 
classical integral (\ref{e.secondtry}) are indeed associated with fine tuning the moduli so as to follow
the trajectories of a lower number of particles\footnote{E.g. for the case of two particles,
there is a direction in moduli
space so that the operator stays on top of one outgoing or ingoing soliton. This noncompact integration 
leads to a divergence associated with the single particle.}. The structural similarity of the divergence
structures of the quantum connected form factor and the classical integral (\ref{e.secondtry})
strongly suggests that there should be a choice of subtraction scheme in (\ref{e.secondtry}) which exactly reproduces
the classical limit of diagonal form factors. The goal of the first part of this paper is indeed to explicitly 
propose such a scheme and thus to provide a classical formula for the connected $n$-particle diagonal form factor
in an arbitrary integrable QFT. This is done in section~3.

In the case of a finite volume system, the moduli space is compact and the integral is finite.
However, exact finite volume multiparticle solutions are exceedingly complicated to construct and
are usually not known explicitly. Despite that, once we allow ourselves to neglect exponential $e^{-mL}$ terms,
we can construct approximate finite volume solutions by gluing together infinite volume solutions.
This has been used in \cite{Bajnok:2014sza} for computing 
the HHL OPE coefficient for a two particle state. Here we give a formulation valid for any number of particles.
Again we have to deal with a moduli space, but now it becomes a quotient of the infinite volume moduli space
by some set of identifications $\Gamma$ which are induced by the gluing procedure. This gluing procedure is
not completely trivial as one has to take into account the classical time delays due to particle scattering.
Using this procedure we may decompose the finite volume expectation value in terms of diagonal infinite volume
form factors and coefficients involving (the classical limit of) Bethe ansatz Jacobian subdeterminants.
This is a very nontrivial consistency check of our proposal for the classical formula for the connected diagonal
form factor. All this is discussed in section~4 of the present paper.

\section{Diagonal form factors and expectation values in infinite volume}

In this section we summarize the definition of diagonal form factors.
We propose formulas for their classical counterparts and check our
ideas on the example of the sine-Gordon theory. 

\subsection{Diagonal form factors }

Form factors are the matrix elements of local operators between asymptotic
(initial or final) states: 
\begin{equation}
\langle p_{m},\dots,p_{1}\vert\mathcal{O}(x,t)\vert p_{1}',\dots,p_{n}'\rangle=
e^{i\Delta Et-i\Delta Px}\langle p_{m},\dots,p_{1}\vert\mathcal{O}\vert p_{1}',
\dots,p_{n}'\rangle\label{eq:O(x,t)FF}
\end{equation}
 In an initial state particles are ordered as $p'_{1}>\dots>p'_{n}$,
while in a final state oppositely. These two types of states are connected
by the multiparticle scattering matrix, which factorizes into the
product of two particle scatterings%
\footnote{We assume that we are either in a theory with one single particle
type, or in a diagonally scattering subsector of a nondiagonal theory,
otherwise, we have to decorate both the states and the scattering
matrix with color labels.%
}:
\begin{equation}
\vert p_{1},\dots,p_{n}\rangle=\prod_{i<j}S(p_{i},p_{j})\vert p_{n},\dots,p_{1}\rangle
\end{equation}
 The two particle scattering matrix satisfies unitarity $S(p_{1},p_{2})S(p_{2},p_{1})=1$.
The adjoint state is denoted by 
$\vert p_{1},\dots,p_{n}\rangle^{\dagger}=\langle p_{n},\dots,p_{1}\vert$
and we choose the following normalization
\begin{equation}
\langle p_{n},\dots,p_{1}\vert p'_{1},\dots,p'_{n}\rangle=\prod_{i=1}^{n}2\pi E(p_{i})
\delta(p_{i}-p'_{i})\label{eq:normalization}
\end{equation}
Both the initial and final states are eigenstates of the conserved
charges including the momentum and the Hamiltonian 
\begin{equation}
P\vert p_{1},\dots,p_{n}\rangle=\sum_{i=1}^{n}p_{i}\vert p_{1},\dots,p_{n}\rangle\quad;\qquad 
H\vert p_{1},\dots,p_{n}\rangle=\sum_{i=1}^{n}E(p_{i})\vert p_{1},\dots,p_{n}\rangle
\end{equation}
As the Hamiltonian generates time, while the momentum space evolution
the space-time dependence of the matrix element can be easily determined
(\ref{eq:O(x,t)FF}), where $\Delta$ denotes the difference of the
quantities on the two sides. In particular, the diagonal matrix element
is independent of the insertion point and depends only on one set
of momenta. This diagonal limit is not well defined, however, due
to disconnected terms. Indeed, let us shift the momenta between the
two sets of rapidities as $p_{i}'=p_{i}+\epsilon_{i}$ and investigate
the $\epsilon_{i}\to0$ limit. Crossing relation \cite{Smirnov:1992vz} allows one to put
a particle with momentum $p$ from the final state into an antiparticle
with momentum $\bar{p}$ in the initial state as
\begin{eqnarray}
\langle p_{n},\dots,p_{2},p_{1}\vert\mathcal{O}\vert p'_{1},p'_{2},\dots p_{n}'\rangle
& = & \langle p_{n},\dots,p_{2}\vert\mathcal{O}\vert\bar{p}_{1},p_{1}',\dots,p_{n}'\rangle+\\
 &  & \langle p_{1}\vert p'_{1}\rangle\langle p_{n},\dots,p_{2}\vert\mathcal{O}\vert p_{2}'
 ,\dots,p_{n}'\rangle+\dots\nonumber 
\end{eqnarray}
where we kept explicitly only the disconnected piece which survives
in the diagonal limit. By crossing all particles and keeping only
the relevant disconnected terms we can express the diagonal matrix
element in terms of the ``elementary'' form factors -having vacuum
in the adjoint state- as 
\begin{eqnarray}
\langle p_{n},...,p_{1}\vert\mathcal{O}\vert p'_{1},\dots p'_{n}\rangle 
& = & \langle0\vert\mathcal{O}\vert\bar{p}_{n},...,\bar{p}_{1},p'_{1},\dots,p'_{n}\rangle\\
 &  & +\sum_{i}\langle p_{i}\vert p'_{i}\rangle\langle0\vert\mathcal{O}\vert\bar{p}_{n},..,
 \hat{\bar{p}}_{i},..,\bar{p}_{1},p'_{1},..,\hat{p}_{i},..,p'_{n}\rangle\nonumber \\
 &  & +\sum_{i,j}\langle p_{i},p_{j}\vert p'_{i},p_{j}'\rangle\langle0\vert\mathcal{O}
 \vert\bar{p}_{n},..,\hat{\bar{p}}_{i},..,\hat{\bar{p}}_{j},..,\bar{p}_{1},p'_{1},..,
 \hat{p}_{i},..,\hat{p}_{j}..,p'_{n}\rangle+\dots\nonumber 
\end{eqnarray}
where terms with hats are absent. In taking the diagonal limit $p_{i}'\to p_{i}$
we face two types of divergences. First, the states are normalized
to delta functions (\ref{eq:normalization}). This can be cured either
by subtracting the disconnected pieces or by putting the system into
a finite volume. The second singularity type comes from taking the
limit in the elementary form factor: 
\begin{equation}
\langle0\vert\mathcal{O}\vert\bar{p}_{n},\dots,\bar{p}_{1},p_{1}+\epsilon_{1},\dots,p_{n}
+\epsilon_{n}\rangle=\frac{\sum_{\{i_{1},\dots,i_{n}\}}a_{i_{1}\dots i_{n}}\epsilon_{i_{1}}
\dots\epsilon_{i_{n}}}{\epsilon_{1}\dots\epsilon_{n}}+\dots
\end{equation}
where we indicated the most singular terms. Clearly the expression
depends on which way we take the diagonal limit. There are two typical
definitions: the symmetric and the connected ones. In this paper we
focus only on the connected evaluation%
\footnote{The other can be easily obtained by the kinematical singularity axiom
of the form factors \cite{Pozsgay:2007gx}.%
}, which is defined as the finite, $\epsilon$-independent, term in
the expansion:
\begin{equation}
F_{n}(p_{1},\dots,p_{n})=n!a_{1\dots n}
\end{equation}
With this definition the diagonal matrix element, what we also call
as the expectation value, can be formally written as: 
\begin{eqnarray}
\langle p_{1},...,p_{n}\vert\mathcal{O}\vert p_{n},\dots p_{1}\rangle 
& = & \sum_{A\subseteq\{1,\dots,n\}}\langle A\vert A\rangle F_{\vert\bar{A}\vert}
\{\bar{A}\}\label{eq:diagffn}\\
 & = & F_{n}+\sum_{i}\langle i\vert i\rangle F_{n-1}\{1,..,\hat{i},..n\}
 +\sum_{i,j}\langle i,j\vert j,i\rangle F_{n-2}\{1,..,\hat{i},..,\hat{j},..,n\}
 +\dots\nonumber 
 \label{diagvsff}
\end{eqnarray}
where $\bar{A}$ is the complement of $A$ i.e. $\bar{A}=\{1,\dots,n\}\setminus A$.
We give a more concrete meaning of this formula by putting the system
into a finite volume and carefully defining the innerproducts of the
states. Alternatively, assuming that we can evaluate the expectation
values, we can express the connected diagonal form factors recursively.
We spell out the details for the $1$ and $2$-particle states: The
1-particle expectation value can be written as

\begin{equation}
\langle p\vert\mathcal{O}\vert p\rangle=F_{1}(p)+\langle p\vert 
p\rangle F_{0}\label{eq:1ptexpvalue}
\end{equation}
or, alternatively, the connected diagonal form factor in terms of
the expectation value reads as 
\begin{equation}
F_{0}=\langle0\vert\mathcal{O}\vert0\rangle\quad;\qquad F_{1}(p)=
\langle p\vert\mathcal{O}\vert p\rangle-\langle p\vert p\rangle\langle0
\vert\mathcal{O}\vert0\rangle\label{eq:1pfffromev}
\end{equation}
The analogous relations for the two particle case are as follows

\begin{equation}
\langle p_{2},p_{1}\vert\mathcal{O}\vert p_{1},p_{2}\rangle=F_{2}(p_{1},p_{2})
+\langle p_{1}\vert p{}_{1}\rangle F_{1}(p_{2})+\langle p_{2}\vert p{}_{2}
\rangle F_{1}(p_{1})+\langle p_{1},p_{2}\vert p_{2},p_{1}\rangle F_{0}
\end{equation}
or alternatively
\begin{eqnarray}
F_{2}(p_{1},p_{2}) & = & \langle p_{2},p_{1}\vert\mathcal{O}\vert p_{1},p_{2}\rangle
-\langle p_{1}\vert p{}_{1}\rangle\langle p_{2}\vert\mathcal{O}\vert p_{2}\rangle
-\langle p_{2}\vert p{}_{2}\rangle\langle p_{1}\vert\mathcal{O}\vert p_{1}\rangle
+\langle p_{1},p_{2}\vert p_{2},p_{1}\rangle\langle0\vert\mathcal{O}\vert0\rangle\nonumber \\
 & = & \langle p_{2},p_{1}\vert\mathcal{O}\vert p_{1},p_{2}\rangle-\langle p_{1}
 \vert p{}_{1}\rangle F_{1}(p_{2})-\langle p_{2}\vert p{}_{2}\rangle F_{1}(p_{1})
 -\langle p_{1},p_{2}\vert p_{2},p_{1}\rangle F_{0}\label{eq:2ptffev}
\end{eqnarray}
We will see analogous relations in the classical limit.

\subsection{Classical limit of diagonal form factors }

In this subsection we propose an expression for the classical limit
of the previously introduced diagonal form factors. In describing
the limit we first note that the diagonal form factor can be thought
of as the regularized quantum average of the operator $\mathcal{O}(\hat{\varphi}(x,t))$
in a given energy-momentum eigenstate. In the classical limit the
operator will be replaced by the function of the field $\mathcal{O}(\varphi(x,t))$,
while the state should correspond to a multiparticle solution with
the same energy and momentum. Finite energy solutions in a classical
integrable theory have multiparticle interpretations: the energy density
is well concentrated around separated segments of straight lines.
They are thought of as trajectories of particles, which interact locally,
only when they get close to each other. Changing the initial location
of a given particle leads to another solution with the same energy.
Consequently, the space of $n$-particle solutions with a given energy
has a moduli space isomorphic to $\mathbb{R}^{n}$. The quantum average
of the operator $\mathcal{O}(\hat{\varphi}(x,t))$ should correspond
in the classical limit to an average of the function $\mathcal{O}(\varphi(x,t))$
over this moduli space. The infinities, however, which appear for
the expectation values in the quantum theory, are present also at
the classical level, thus we need to introduce proper subtractions.
Performing these subtractions we define a localized function, which
we integrate over the moduli space of the classical solutions with
a given energy. As the strong coupling limit of the HHL solutions
can be mapped by the Pohlmeyer reduction to relativistic scattering
theories we focus in this section on relativistic theories. We analyze
the infinite volume multiparticle solutions first and then focus on
the corresponding form factors. Sometimes it is useful to have explicit
solutions in mind. For this reason we provide explicit formulas for
the sine-Gordon theory which is defined by the Lagrangian 
\begin{equation}
\mathcal{L}=\frac{1}{2}(\partial\varphi)^{2}-\frac{m^{2}}{\beta^{2}}(1-\cos\beta\varphi)
\end{equation}

\subsubsection{Classical solutions and their moduli space}

We consider an integrable classical field theory, which admits finite
energy localized solutions allowing for multiparticle interpretation.
We introduce the moduli space of these solutions by proceeding in
the particle number.

\paragraph*{Vacuum}

The vacuum solution is a translational invariant --constant-- solution
of the equation of motion, which we denote by $\varphi_{0}$. Its
moduli space is discrete and usually consists of one single point.
In the sine-Gordon case this point is $\varphi_{0}=0\equiv\pm\frac{2\pi}{\beta}$.

\paragraph*{1-particle}

The simplest 1-particle solution is the static solution%
\footnote{We introduced a mass parameter $m$ to make $x$ dimensionless. %
}, $\varphi_{st}(mx)$. The energy density of this solution, $\epsilon[\phi_{st}(mx)]$,
is localized sharply around one point, which we choose to be the origin,
$x=0$. Shifting this point spans the moduli space of the static solutions.
Each solution can be interpreted as a standing particle. 

The moving 1-particle solution can be obtained by boosting the static
solution: 
\begin{equation}
\varphi_{st}(m\cosh\theta x-m\sinh\theta(t-t_{1}))=\varphi_{st}(Ex-p(t-t_{1}))
=\varphi_{st}(E(x-x_{1})-pt)
\end{equation}
By introducing the dimensionless variable $y=Ex-pt$ we can write
the moving solution in the form 
\begin{equation}
\varphi_{1}(x,t;y_{1})\equiv\varphi_{st}(y-y_{1})
\end{equation}
 Due to translational invariance the shifted solution is also a solution
and we parametrize the moduli space of the classical 1-particle solutions
-- of a given momentum -- by $y_{1}\in\mathbb{R}$, which is
$y_1=Ex_{1}$. We choose the sign of $y_{1}$, such that
$y_{1}\to\pm\infty$ shifts the particle's trajectory to $\pm\infty$.
We assume that the theory has no internal symmetry, so the translation
$y_{1}$, is the only continuous parameter of the moduli space. This
moduli space is denoted by $\mathcal{M}^{1}=\mathbb{R}$. The 1-particle
solution with a given momentum $p$ can be considered as a function
of the space-time coordinates and the moduli parameter $y_{1}$ and
we denoted this function by $\varphi_{1}(x,t;y_{1})$, i.e. we do
not write out explicitly the momentum dependence. As the energy density
of the solution is concentrated around the zero of the argument of
the static solution, we can think of this solution in terms of a particle's
trajectory:
\begin{equation}
x(t)=v(t-t_{1})=vt+x_{1}\quad;\qquad v=\tanh\theta
\end{equation}

In the sine-Gordon theory the solutions can be most conveniently expressed
in terms of $\tan\frac{\beta\varphi}{4}$. In particular, the 1-particle
solution, $\varphi_{1}(x,t;y_{1})\equiv\varphi_{1}$, reads as 
\begin{equation}
e_{1}\equiv\tan\frac{\beta\varphi_{1}}{4}=e^{m\cosh\theta_{1}(x-x_1)-
m\sinh\theta_{1}t}=e^{y-y_{1}}
\end{equation}
It interpolates between $0$ and $\frac{2\pi}{\beta}$ and is called
the soliton. Anti-solitons can be described either by $-e_{1}$ or
by $e_{1}^{-1}$. Actually $-e_{1}^{-1}$ is the soliton again.

\paragraph*{2-particle}

\begin{figure}
\begin{centering}
\includegraphics[width=5cm]{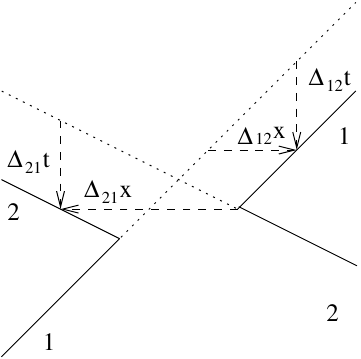}
\par\end{centering}
\protect\caption{Relativistic particle scattering process, in which particle $1$
comes from the left and after scattering on particle $2$ coming from
the right it experiences a $\Delta_{12}x$ space-displacement and $\Delta_{12}t$
time-delays. Particle $2$ has the analogous $\Delta_{21}x$ space-displacement
and $\Delta_{21}t$ time delays. \label{2ptscat}}
\end{figure}

The 2-particle solution with momenta $p_{1}$ and $p_{2}$ denoted
by $\varphi_{2}(x,t;y_{1},y_{2})$ generalizes the 1-particles solution
as follows: the moduli space, $\mathcal{M}^{2}=\mathbb{R}^{2}$, has
two parameters $y_{1}$ and $y_{2}$, which are the respective shifts
in the particles' trajectories, $y_{i}=E_{i}x_{i}$, such that $y_{i}\to\infty$
shifts particle $i$ to $+\infty$. Upto a localized interaction domain,
the solution is the composition of two 1-particle solutions. These
1-particle solutions, however are not the same before and after their
interaction: there is a space displacement and a time delay. Focusing
on the energy density we can interpret the 2-particle solution in
terms of a collision process as follows. The particles travel freely
\begin{equation}
x_{1}(t)=v_{1}t+x_{1}^{-}=v_{1}(t-t_{1}^{-})\quad;\qquad x_{2}(t)=
v_{2}t+x_{2}^{-}=v_{2}(t-t_{2}^{-})
\end{equation}
before they interact, say at time $t=0$. After the interaction they
travel freely again as
\begin{equation}
x_{1}(t)=v_{1}t+x_{1}^{+}=v_{1}(t-t_{1}^{+})\quad;\qquad x_{2}(t)=v_{2}t+x_{2}^{+}=v_{2}(t-t_{2}^{+})
\end{equation}
The result of the interaction is the experienced time delays or space
displacements: 
\begin{equation}
\Delta_{12}t=t_{1}^{+}-t_{1}^{-}\quad;\qquad\Delta_{21}t=t_{2}^{+}-t_{2}^{-}\quad;
\qquad\Delta_{12}x=x_{1}^{+}-x_{1}^{-}\quad;\qquad\Delta_{21}x=x_{2}^{+}-x_{2}^{-}
\end{equation}
We show the scattering process on the schematic Figure \ref{2ptscat},
where, to be specific, we assumed that $p_{1}>0,p_{2}<0$ such that
the space displacements have opposite signs: $\Delta_{12}x=-v_{1}\Delta_{12}t>0$
and $\Delta_{21}x=-v_{2}\Delta_{21}t<0$. These displacements are
not the same for the two particles but can be related via the free
movement of the center of energy:
\begin{equation}
\frac{E_{1}x_{1}+E_{2}x_{2}}{E_{1}+E_{2}}=\frac{E_{1}v_{1}+E_{2}v_{2}}{E_{1}+E_{2}}t
+\frac{E_{1}x_{1}^{\pm}+E_{2}x_{2}^{\pm}}{E_{1}+E_{2}}
\end{equation}
Expressing this motion in terms of the quantities before and after
the interaction leads to the relations 
\begin{equation}
E_{1}\Delta_{12}x+E_{2}\Delta_{21}x=0\qquad;\qquad p_{1}\Delta_{12}t+p_{2}\Delta_{21}t=
0\label{eq:D12+D21}
\end{equation}
where we used that $Ev=p$. As the 2-particle solution is the classical
limit of a scattering process it is interesting to relate the appearing
quantities to the S-matrix. The energy derivative of the phase shift
of the S-matrix is related in the semiclassical limit to the time
delay as \cite{Jackiw:1975im}:
\begin{equation}
(\partial_{E_{1}}p_{1})\partial_{p_{1}}\delta(p_{1},p_{2})\to\Delta_{12}t\qquad;
\qquad S=e^{i\delta(p_{1},p_{2})}
\end{equation}
In particular, we can relate the time delays and space displacements
to the classical limit, $\phi_{12}^{c}$, of the quantity $\phi(p_{1},p_{2})=
E_{1}\partial_{p_{1}}\delta(p_{1},p_{2})$
as 
\begin{equation}
\phi(p_{1},p_{2})=E_{1}\partial_{p_{1}}\delta(p_{1},p_{2})\to\phi_{12}^{c}=
E_{1}\frac{\partial E_{1}}{\partial p_{1}}\Delta_{12}t=p_{1}\Delta_{12}t=
-E_{1}\Delta_{12}x=-\Delta_{12}y
\end{equation}
i.e. the shift in the moduli space is nothing but the classical limit
of $-\phi(p_{1},p_{2})$. The shift for the other particle is 
\begin{equation}
-\phi(p_{2},p_{1})=E_{2}\partial_{p_{2}}\delta(p_{1},p_{2})\to-\phi_{21}^{c}=
E_{2}\frac{\partial E_{2}}{\partial p_{2}}\Delta_{21}t=p_{2}\Delta_{21}t=
-E_{2}\Delta_{21}x=-\Delta_{21}y
\end{equation}
We can see from (\ref{eq:D12+D21}) that the shifts in the moduli
parameters sum up to zero: $\Delta_{12}y+\Delta_{21}y=0$. This motivates
us to work with this moduli parameter and not with the space displacements
or time delays. 

In the sine-Gordon theory the 2-soliton solution,
$\varphi_{2}(x,t,y_{1},y_{2})\equiv\varphi_{2}$,
can be written in terms of the two 1-soliton solutions as 
\cite{Bryan:1988}
\begin{equation}
\tan\frac{\beta\varphi_{2}}{4}\equiv e_{12}=\frac{e_{1}+e_{2}}{1-u_{12}^{2}e_{1}e_{2}}
\quad;\qquad u_{12}=\tanh\frac{\theta_{1}-\theta_{2}}{2}\quad;\qquad e_{i}=
e^{m\cosh\theta_{i}x-m\sinh\theta_{i}t-y_{i}}
\end{equation}
This solution, except for some local interaction domain, can be considered
as two non-interacting one soliton solutions. The effect of the interaction
is that the solitons experience some time delays. To calculate these
time delays we analyze the solutions in the asymptotic limits. As
the energy density is proportional to $\frac{1}{(e_{12}+e_{12}^{-1})^{2}}$
the nontrivial contributions come from the domains when $e_{i}$ is
not close either to $0$ or to $\infty$. These are the places where
the solitons are localized and agree with the zero of the exponent
of $e_{i}$: $E_{i}x-p_{i}t-y_{i}=0$. Analyzing the $t\to-\infty$
limit we can see two nontrivial domains contributing. For $x<0$ the
quantity $e_{2}$ vanishes, while for $x>0$ the other $e_{1}$ goes
to infinity leading to 
\begin{equation}
e_{12}=\begin{cases}
\begin{array}{c}
e_{1}\qquad\mbox{for}\quad x<0\\
-\frac{1}{u_{12}^{2}e_{2}}\quad\mbox{for}\quad x>0
\end{array}\end{cases}
\end{equation}
 We can reparametrize the $x>0$ soliton as 
\begin{equation}
-\frac{1}{u_{12}^{2}e_{2}}\to u_{12}^{2}e_{2}=e^{m\cosh\theta_{2}x-m\sinh\theta_{2}t
+\phi_{12}^{c}}\quad;\qquad\phi_{12}^{c}=\log(\tanh^{2}(\frac{\theta_{1}-\theta_{2}}{2}))
\end{equation}
In the $t\to\infty$ we found 
\begin{equation}
e_{12}=\begin{cases}
\begin{array}{c}
e_{2}\qquad\mbox{for}\quad x<0\\
-\frac{1}{u_{12}^{2}e_{1}}\quad\mbox{for}\quad x>0
\end{array}\end{cases}
\end{equation}
Parametrizing the particles' trajectories before and after the collision
as $E_{i}x-p_{i}(t-t_{i}^{\pm})=0$ we can read off that before the
collision $ $$t_{1}^{-}=0$ and $t_{1}^{+}=p_{1}^{-1}\phi_{12}^{c}$,
while after the collision $t_2^+=0$ and $t_{2}^{-}=p_{2}^{-1}\phi_{12}^{c}$. These
lead to the following time delays:
\begin{equation}
\Delta_{12}t=t_{1}^{+}-t_{1}^{-}=\frac{\phi_{12}^{c}}{m\sinh\theta_{1}}=
\frac{\phi_{12}^{c}}{p_{1}}\quad;\qquad\Delta_{21}t=t_{2}^{+}-t_{2}^{-}=
0-\frac{\phi_{12}^{c}}{m\sinh\theta_{2}}=-\frac{\phi_{12}^{c}}{p_{2}}
\end{equation}
Clearly the relation $p_{1}\Delta_{12}t+p_{2}\Delta_{21}t=0$ is satisfied.

\paragraph*{n-particle}

The $n$ particle solution with momenta $p_{1},\dots,p_{n}$ denoted
by $\varphi_{n}(x,t;y_{1},\dots,y_{n})$ depends on the space-time
coordinates and on the moduli parameters $y_{i}\in\mathcal{M}^{n}=\mathbb{R}^{n}$,
which are the respective translations of each individual particles.
By shifting the leftmost particle to $y_{1}\to-\infty$ the $n$ particle
solution reduces to the $n-1$ particle solution: 
$\varphi_{n}(x,t;\infty,y_{2}\dots,y_{n})=\varphi_{n-1}(x,t;y_{2},\dots,y_{n})$.
By shifting the same particle to $y_{1}\to\infty$ it scatters on
each particle and suffers $\sum_{j=2}^{n}\Delta_{1j}y$ displacements.
Additionally, it shifts the other particles by $\Delta_{j1}y$ leading
to the solution 
$\varphi_{n}(x,t;\infty,y_{2},\dots,y_{n})=\varphi_{n-1}(x,t;y_{2}+
\Delta_{21}y_{2},\dots,y_{n}+\Delta_{n1}y)$.
In general, the $n$ particle solution reduces to the $n-k$ particle
solution, whenever the other $k$ particles are translated to infinity. 

In the sine-Gordon theory the n-soliton solution,
$\varphi_{n}(x,t;y_{1},\dots,y_{n})\equiv\varphi_{n}$,
can be written as \cite{Bryan:1988}
\begin{equation}
\tan\frac{\beta\varphi_{n}}{4}=\frac{\Im m(\tau)}{\Re e(\tau)}\quad;\qquad\tau=\sum_{\mu_{j}=
\{0,1\}}\prod_{j=1}^{n}(ie_{j})^{\mu_{j}}\prod_{i<j}u_{ij}^{2\mu_{i}\mu_{j}}
\end{equation}
where 
\begin{equation}
e_{i}=e^{m\cosh\theta_{i}x-m\sinh\theta_{i}t-y_{i}}\quad;\qquad u_{ij}=
\tanh\frac{\theta_{i}-\theta_{j}}{2}
\end{equation}
The classical time delay of the $i^{th}$ particle after passing through
the $j^{th}$ particle is independent of the other particles and reads
as 
\begin{equation}
\Delta_{ij}t=\frac{\phi_{ij}^{c}}{p_{i}}\qquad;\qquad\phi_{ij}^{c}=
\log\tanh^{2}\frac{\theta_{i}-\theta_{j}}{2}
\end{equation}

\subsubsection{Classical form factors}

As we mentioned before the quantum average of the operator $\mathcal{O}(\hat{\varphi}(x,t))$
should correspond in the classical limit to an average of the function
$\mathcal{O}(\varphi(x,t))$ over the moduli space of classical solutions.
Since the infinities which appear for the expectation values in the
quantum theory are present also at the classical level we develop
proper subtraction procedure. We proceed in the particle number. For
reference we present the form factors of the trace of the energy-momentum
tensor in the sine-Gordon theory 
\begin{equation}
\Theta^{c}(\varphi)=\frac{m^{2}}{\beta^{2}}(1-\cos\beta\varphi)=
\frac{8m^{2}}{\beta^{2}\left(\tan\frac{\beta\varphi}{4}+\cot\frac{\beta\varphi}{4}\right)^{2}}
\end{equation}

\paragraph*{Vacuum}

The classical limit of the vacuum is the constant classical vacuum
solution $\varphi_{0}$ and the classical limit of the vacuum expectation
value of the operator $\mathcal{O}(\hat{\varphi})$ is simply its
value $\mathcal{O}[\varphi_{0}]$. If there are many vacua then the
expression might depend on which vacuum we evaluate the operator. 

In the sine-Gordon theory $\Theta(\varphi)$ is vanishing on the vacuum
$\varphi_{0}=0\equiv\frac{2\pi}{\beta}$.

\paragraph*{1-particle}

The classical limit of a 1-particle asymptotic state is the moving
1-particle solution. Its moduli space is $\mathcal{M}_{1}=\mathbb{R}$
and the classical analogue of the quantum expectation value should
be the average for the moduli parameter $y_{1}$: 
\begin{equation}
\langle p\vert\mathcal{O}\vert p\rangle^{c}\to\int_{\mathcal{M}^{1}}dy_{1}\,
\mathcal{O}(\varphi_{1}(x,t,y_{1}))
\end{equation}
Similarly, however, to the quantum case (\ref{eq:1ptexpvalue}) the
expression is divergent if the operator has a vacuum expectation value.
Analogy with the quantum case suggests to define the diagonal form
factor after a proper subtraction (\ref{eq:1pfffromev}): We define
the classical 1-particle diagonal form factor of the operator $\mathcal{O}(\varphi)$
to be the integral 
\begin{equation}
F_{1}^{c}=\int_{-\infty}^{\infty}dy_{1}\,\left\{ \mathcal{O}(\varphi_{1}(x,t,y_{1}))
-\mathcal{O}(\varphi_{0})\right\} \label{eq:1ptclff}
\end{equation}
As the $1$-particle solution agrees with the vacuum solution away
from the trajectory of the particle the function 
$\mathcal{O}(\varphi_{1})-\mathcal{O}(\varphi_{0})$
is well localized. Consequently, the integral has a finite support
and gives a finite result. As the moduli parameter $y_{1}$ shifts
the classical solution (both in space and in time) the integral is
actually independent of the space-time coordinates $(x,t)$. This
fits very well to the picture of being the classical limit of the
quantum diagonal form factor, which is also space-time independent.
As this will be true also for multiparticle form factors we put $x=t=0$
and omit to write out the space-time coordinates 
$\varphi_{n}(y_{1},\dots,y_{n})\equiv\varphi_{n}(0,0;y_{1},\dots,y_{n})$.
To further simplify our formulas we analyze operators without vacuum
expectation values. This can be easily arranged by redefining the
operators as $\mathcal{O}(\varphi)\to(\mathcal{O}(\varphi)-\mathcal{O}(\varphi_{0}))$.
These newly defined observables are then localized where the particles
are localized. In particular, the 1-particle integral (\ref{eq:1ptclff})
collects its contribution from a small domain around $y_{1}=0$, which
is indicated with a black dot in the moduli space. 

\begin{figure}
\begin{centering}
\includegraphics[width=5cm]{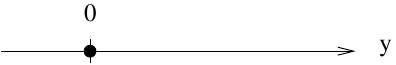}
\par\end{centering}
\protect\caption{One particle moduli space $\mathcal{M}_{1}=\mathbb{R}$. Black dot
indicates the point, whose neighbourhood contributes to the $1$-particle
form factor.}
\end{figure}

In the sine-Gordon theory the one-particle connected diagonal form
factor of $\Theta$ is 
\begin{equation}
F_{1}^{\Theta}(\theta)=\frac{1}{4}(F_{1}^{T_{00}}-F_{1}^{T_{11}})=
\frac{M^{2}}{4}(\cosh^{2}\theta-\sinh^{2}\theta)=\frac{M^{2}}{4}
\end{equation}
where $M$ is the soliton mass. Let us calculate the classical form
factor from (\ref{eq:1ptclff}): 
\begin{equation}
F_{1}^{c}=\int_{-\infty}^{\infty}dy_{1}\,\frac{m^{2}}{\beta^{2}}(1-\cos\beta\varphi_{1})=
\frac{8m^{2}}{\beta^{2}}\int_{0}^{\infty}\frac{de_{1}}{e_{1}}\frac{1}{(e_{1}+e_{1}^{-1})^{2}}=
\frac{4m^{2}}{\beta^{2}}
\end{equation}
which is consistent with the quantum formula as the classical limit
of the soliton mass is $M^{c}=\frac{4m}{\beta}$.

\paragraph{2-particle}

The moduli space of the two particle solution, $\mathcal{M}_{2}=\mathbb{R}^{2}$,
contains separate shifts in each particle's locations $y_{1}$ and
$y_{2}$. The classical analogue of the quantum average should correspond
to the integral

\begin{equation}
\langle p_{2},p_{1}\vert\mathcal{O}\vert p_{1},p_{2}\rangle^{c}\to\int_{\mathcal{M}^{2}}
\, dy_{1}dy_{2}\mathcal{O}(\varphi_{2}(y_{1},y_{2}))\label{eq:2ptclav}
\end{equation}
However, as the quantum formula (\ref{eq:2ptffev}) suggests the integral
is infinite whenever the one particle form factor is nonzero. Indeed,
for operators without vacuum expectation value, the contributions
come from the trajectories of the particles, which form the scattering
process on Figure \ref{2ptscat}. Let us analyze this two particle
scattering picture and insert the operator at the origin to see the
effects of the various shifts in $y_{i}$. Technically it is simpler
to draw the particle trajectories unchanged and shift the operator
in the opposite way, see Figure \ref{2ptopins}. The characteristic
quantity in the process is the space-time displacements, which translate
to the moduli parameter as 
$\Delta_{12}y=E_{1}\Delta_{12}x=-\phi_{12}^{c}=-\Delta_{21}y=-E_{2}\Delta_{21}x$.

\begin{center}
\begin{figure}
\begin{centering}
\includegraphics[width=4cm]{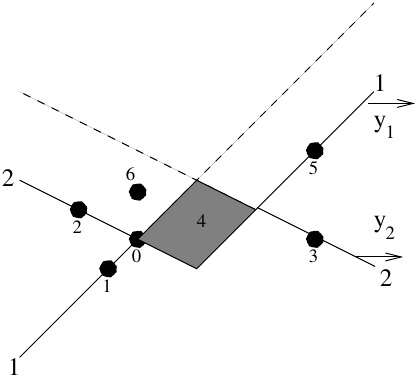}
\par\end{centering}

\protect\caption{The insertion of the operator as compared to the two particle scattering
solutions. Special configurations for the operator are labeled from
$0$ to $6$. $0$ labels the location of the operator for $y_{1}=0$
and $y_{2}=0$ and it is assumed that we collect contributions in
this case. 1: $y_{2}>0$ only contribution from particle 1 at $y_{1}=0$.
2: $y_{1}>0$ only contribution from particle 2 at $y_{2}=0$. 3:
$y_{1}<\phi_{12}^{c}$ only contribution from particle 2 at $y_{2}=\phi_{12}^{c}$.
4: $\phi_{12}^{c}<y_{1}<0$ and $\phi_{12}^{c}<y_{2}<0$ two-particle
contribution. 5: $y_{2}<\phi_{12}^{c}$ only contribution from particle
1 at $y_{1}=\phi_{12}^{c}$. 6: generic point, not mentioned above:
no contribution at all. \label{2ptopins}}
\end{figure}

\par\end{center}

\begin{figure}
\begin{centering}
\includegraphics[width=6cm]{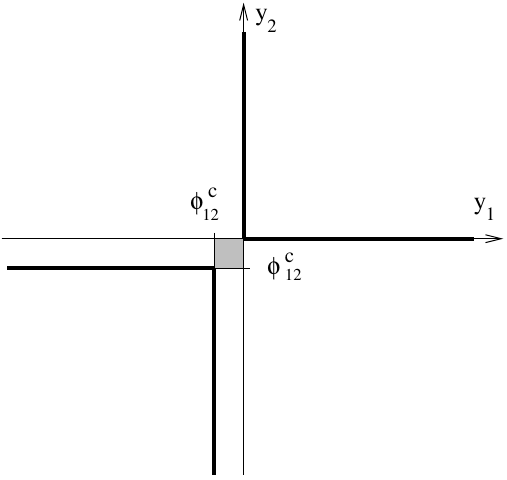}~~~~~~~~~\includegraphics[width=6cm]{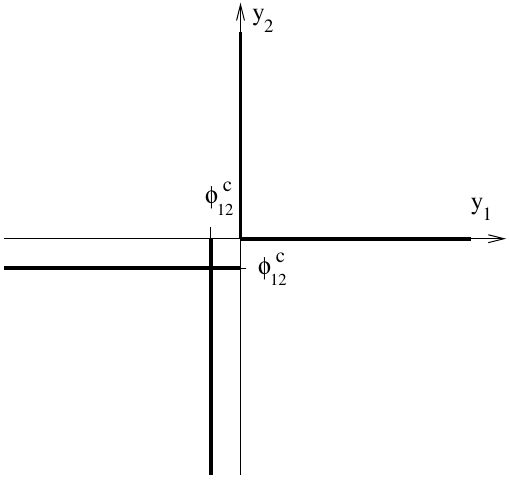}
\par\end{centering}

\protect\caption{Domains in the moduli space $\mathcal{M}_{2}$ where the function
$\mathcal{O}(\varphi_{12}(y_{1},y_{2}))$ takes non-vanishing contributions
are indicated on the left. Subtracted one particle contributions are
indicated on the right. \label{2pffmodsub}}
\end{figure}
The translation of Figure \ref{2ptopins} into the moduli space tells
the domain where the particles are located or, equivalently, the domain
where the function $\mathcal{O}(\varphi_{12}(y_{1},y_{2}))$ is non-vanishing
and the integral (\ref{eq:2ptclav}) collects its contributions from.
See the left of Figure \ref{2pffmodsub}. Near the 1-particle lines
the other particle is far away and the solution can be approximated
with a 1-particle solution, which depends only on one moduli parameter.
The integral for the other moduli parameter will then give infinite
contribution. To define a finite quantity we have to subtract the
contributions of the infinite one particle lines. These one particle
lines are not the same before and after the interactions, i.e. they
are shifted by $\Delta_{12}y=-\phi_{12}^{c}$. The interaction domain
is localized within a square of size $\Delta_{12}y$ and in subtracting
the one particle lines we have an ambiguity in choosing the end and
the start of the shifted semi-infinite lines. Different choices lead
to different form factors and we present here only the one, which
corresponds to the classical limit of the connected form factors,
see the right of Figure \ref{2pffmodsub}. From the subtraction point
of view we consider the interaction to be point like at $y_{1}=y_{2}=0$.
For $y_{1}<0$ we shift particle $1$ to $-\infty$, while for $y_{1}>0$
we shift it to $+\infty$ and subtract the obtained contributions.
We repeat the same for particle $2$ and arrive at the definition
of the classical two particle diagonal form factor:

\begin{eqnarray}
F_{2}^{c}(p_{1},p_{2})=\int_{-\infty}^{\infty}dy_{1}\int_{-\infty}^{\infty}dy_{2}
\biggl[\mathcal{O}[\varphi_{2}(y_{1},y_{2})]\,-\Theta(-y_{1})\mathcal{O}
[\varphi_{2}(-\infty,y_{2})]-\Theta(y_{1})\mathcal{O}[\varphi_{2}(\infty,y_{2})]\:\:\\
-\Theta(-y_{2})\mathcal{O}[\varphi_{2}(y_{1},-\infty)]-\Theta(y_{2})
\mathcal{O}[\varphi_{2}(y_{1},\infty)]\biggr]\nonumber 
\end{eqnarray}
This integrand is localized in both moduli parameter in a finite domain
around the origin denoted by the shadowed region on the left of Figure
\ref{2pffmodsub}, i.e. on $\phi_{12}^{c}<y_{1}<0$ and $\phi_{12}^{c}<y_{2}<0$. 

In the sine-Gordon theory the 2-particle connected form factor of
$\Theta$ is 
\begin{equation}
F_{2}^{\Theta}(\theta_{1}-\theta_{2})=\frac{1}{4}(F_{2}^{T_{00}}-F_{2}^{T_{11}})=
\frac{M^{2}}{4}2\phi_{12}(\cosh\theta_{1}\cosh\theta_{2}-\sinh\theta_{1}\sinh\theta_{2})=
\frac{M^{2}}{2}\phi_{12}\cosh(\theta_{1}-\theta_{2})
\end{equation}
We can compare the classical limit of this expression with our definition,
which reads as 
\begin{eqnarray}
F_{2}^{c}(\theta_{1},\theta_{2})=\frac{8m^{2}}{\beta^{2}}\int_{-\infty}^{\infty}dy_{1}
\int_{-\infty}^{\infty}dy_{2}\biggl[\frac{1}{(e_{12}+e_{12}^{-1})^{2}}\,-\Theta(-y_{1})
\frac{1}{(u_{12}^{2}e_{2}+(u_{12}^{2}e_{2})^{-1})^{2}}\,\,\,\,\,\,\,\,\,\,\,\,\,\,\,\,\,\,\,\,\,\\
\,-\Theta(y_{1})\frac{1}{(e_{2}+e_{2}^{-1})^{2}}-\Theta(-y_{2})
\frac{1}{(u_{12}^{2}e_{2}+(u_{12}^{2}e_{2})^{-1})^{2}}-\Theta(y_{2})
\frac{1}{(e_{1}+e_{1}^{-1})^{2}}\biggr]\nonumber 
\end{eqnarray}
Alternatively we can change the integration variables for $e_{1}$
and $e_{2}$: 
\begin{eqnarray}
F_{2}^{c}(\theta_{1},\theta_{2})=\frac{8m^{2}}{\beta^{2}}\int_{0}^{\infty}\frac{de_{1}}{e_{1}}
\int_{0}^{\infty}\frac{de_{2}}{e_{2}}\biggl[\frac{1}{(e_{12}+e_{12}^{-1})^{2}}\,-
\Theta(1-e_{1})\frac{1}{(u_{12}^{2}e_{2}+(u_{12}^{2}e_{2})^{-1})^{2}}
\:\:\,\,\,\,\,\,\,\,\,\,\,\,\,\,\,\,\,\,\,\,\,\,\,\,\,\,\,\,\,\,\,\,\,\,\\
-\Theta(e_{1}-1)\frac{1}{(e_{2}+e_{2}^{-1})^{2}}-\Theta(1-e_{2})
\frac{1}{(u_{12}^{2}e_{2}+(u_{12}^{2}e_{2})^{-1})^{2}}-\Theta(e_{2}-1)
\frac{1}{(e_{1}+e_{1}^{-1})^{2}}\biggr]\nonumber 
\end{eqnarray}
Since $e_{12}=\frac{e_{1}+e_{2}}{1-u_{12}^{2}e_{1}e_{2}}$ the integral
depends only on $u_{12}^{2}$. We managed to perform this integral
and obtained 
\begin{equation}
F_{2}^{c}(\theta_{1},\theta_{2})=-\frac{4m^{2}}{\beta^{2}}\frac{u_{12}^{2}+1}{u_{12}^{2}-1}
\log u_{12}^{4}=\frac{8m^{2}}{\beta^{2}}\cosh(\theta_{1}-\theta_{2})\log\tanh^{2}\frac{\theta_{1}-\theta_{2}}{2}
\end{equation}
which is the classical limit of the connected diagonal form factor. 

For the application of the HHL three point functions we calculate
the classical form factors of the operators 
\begin{equation}
\mathcal{O}_{k}(\varphi)=e^{ik\beta\varphi}-1
\end{equation}
in the sine-Gordon theory. As these operators do not have any vacuum
expectation value the 1-particle form factor is obtained as 
\begin{equation}
F_{1}^{\mathcal{O}_{k}}=\int_{-\infty}^{\infty}dy_{1}\,(e^{ik\beta\varphi_{1}}-1)=
\int_{0}^{\infty}\frac{de_{1}}{e_{1}}\left\{ \left(\frac{2i-e_{1}+e_{1}^{-1}}{e_{1}+e_{1}^{-1}}
\right)^{2k}-1\right\} 
\end{equation}
Performing the integral we found 
\begin{equation}
F_{1}^{\mathcal{O}_{k}}=\left\{ -4,-\frac{16}{3},-\frac{92}{15},-\frac{704}{105}\right\} 
\quad;\qquad k=1,2,3,4
\end{equation}
for the first few cases. In the following we focus on the two particle
form factors and evaluate the general formula
\begin{eqnarray}
F_{2,c}^{\mathcal{O}_{k}}=\int_{-\infty}^{\infty}dy_{1}\int_{-\infty}^{\infty}dy_{2}
\biggl[\mathcal{O}_{k}[\varphi_{2}(y_{1},y_{2})]-\Theta(-y_{1})\mathcal{O}_{k}[\varphi_{2}
(-\infty,y_{2})]\,\,\,\,\,\,\,\,\,\,\,\,\,\,\,\,\,\,\,\,\,\,\,\,\,\,\,\,\,\,\\
-\Theta(y_{1})\mathcal{O}_{k}[\varphi_{2}(\infty,y_{2})]-\Theta(-y_{2})\mathcal{O}_{k}
[\varphi_{2}(y_{1},-\infty)]-\Theta(y_{2})\mathcal{O}_{k}[\varphi_{2}(y_{1},\infty)]\nonumber 
\end{eqnarray}
First, using the definition of $e_{12}$, we can rewrite the operator
as 
\begin{equation}
\mathcal{O}_{k}(\varphi)=e^{ik\beta\varphi}-1=\left(\frac{2i-e_{12}+e_{12}^{-1}}{e_{12}+e_{12}^{-1}}
\right)^{2k}-1
\end{equation}
The integrand can alternatively be reformulated as 
\begin{eqnarray}
\mathcal{O}_{k}(y_{1},y_{2})-\frac{1}{1+e^{y_{1}}}\mathcal{O}_{k}(-\infty,y_{2})
-\frac{1}{1+e^{y_{2}}}\mathcal{O}_{k}(y_{1},-\infty)\nonumber \\
-\frac{e^{y_{1}}}{1+e^{y_{1}}}\mathcal{O}_{k}(\infty,y_{2})
-\frac{e^{y_{2}}}{1+e^{y_{2}}}\mathcal{O}_{k}(y_{1},\infty)
\end{eqnarray}
since the difference integrates to zero. In the following we change
variables from $(y_{1},y_{2})$ to $(e_{1},e_{2})$. Clearly the integral
depends only on $u_{12}=\tanh\frac{\theta_{1}-\theta_{2}}{2}$, what we
abbreviate by $u$ in the following. We performed the two integrals
one after the other and obtained the following result: 
\begin{eqnarray}
F_{2}^{\mathcal{O}_{1}}(\theta_{1},\theta_{2})&=& \frac{16\left(u^{2}+1\right)}{u^{2}-1}\log u \nonumber \\
F_{2}^{\mathcal{O}_{2}}(\theta_{1},\theta_{2})&=& \frac{64\left(u^{2}+1\right)^{3}}{3\left(u^{2}-1\right)^{3}} \log u
- \frac{256u^{2}}{3\left(u^{2}-1\right)^{2}} \nonumber \\
F_{2}^{\mathcal{O}_{3}}(\theta_{1},\theta_{2})&=& \frac{16\left(23u^{10}+155u^{8}+590u^{6}+590u^{4}+155u^{2}+23\right)}{15\left(u^{2}-1\right)^{5}}\log u
-\frac{512u^{2}\left(3u^{4}+2u^{2}+3\right)}{5\left(u^{2}-1\right)^{4}}\nonumber \\
F_{2}^{\mathcal{O}_{4}}(\theta_{1},\theta_{2})&=&\frac{256\left(u^{2}+1\right)^{3}\left(11u^{8}+100u^{6}+738u^{4}+100u^{2}+11\right)}{105\left(u^{2}-1\right)^{7}} 
\log u \nonumber \\
&&\hspace{1cm}-\frac{1024u^{2}\left(71u^{8}+180u^{6}+458u^{4}+180u^{2}+71\right)}{105\left(u^{2}-1\right)^{6}}
\end{eqnarray}
Note that the rational part ensures a regular $u=1$ behaviour.
The formulas for higher $k$ get heavy after this point and it would be
nice to find a compact expression for them. 

\paragraph*{n-particle}

In the case of $n$-particles the quantum average should go to the
classical moduli average: 

\begin{equation}
\langle p_{n},\dots,p_{1}\vert\mathcal{O}\vert p_{1},\dots,p_{n}\rangle^{c}\to\int_{\mathcal{M}^{n}}\, 
dy_{1}\dots dy_{n}\mathcal{O}(\varphi_{n}(y_{1},\dots,y_{n}))\label{eq:nptclav}
\end{equation}
To regulate this expression we have to subtract successively the lower
particle number contributions in the spirit of (\ref{eq:diagffn}).
The subtraction, which corresponds to the classical limit of the diagonal
connected form factor reads as: 
\begin{eqnarray}
F_{n}^{c}(p_{1},\dots,p_{n}) & = & \prod_{i}\int_{-\infty}^{\infty}dy_{i}\,\biggl\{\mathcal{O}[\varphi_{n}(y_{1}
,\dots,y_{n})]-\sum_{i,\epsilon_{i}}\Theta(\epsilon_{i}y_{i})\mathcal{O}[\varphi_{n}(y_{1},\dots,
\epsilon_{i}\infty,\dots,y_{n})]\nonumber \\
 &  & +\sum_{i,j,\epsilon_{i},\epsilon_{j}}\Theta(\epsilon_{i}y_{i},\epsilon_{j}y_{j})\mathcal{O}
 [\varphi_{n}(y_{1},\dots,\epsilon_{i}\infty,\dots,\epsilon_{j}\infty,\dots,y_{n})]+\dots\nonumber \\
 &  & +(-1)^{k}\sum_{\{i_{k},\epsilon_{k}\}}\Theta(\epsilon_{i_{1}}y_{i_{1}},\dots,\epsilon_{i_{k}}y_{i_{k}})
 \mathcal{O}[\varphi_{n}(y_{1},\dots,\epsilon_{i_{1}}\infty,\dots,\epsilon_{i_{k}}\infty,\dots,y_{n})]
 +\dots\biggr\}\nonumber \\
 & \equiv & \prod_{i}\int_{-\infty}^{\infty}dy_{i}\,\mathcal{O}[\varphi_{n}(y_{1},\dots,y_{n})]_{c}
\end{eqnarray}
where $\Theta(\epsilon_{i_{1}}y_{i_{1}},\dots,\epsilon_{i_{k}}y_{i_{k}})=\prod_{j=1}^{k}
\Theta(\epsilon_{i_{j}}y_{i_{j}})$.
We can also express each term in terms of the lower order connected
terms. This unifies the signs as:
\begin{eqnarray}
F_{n}^{c}(p_{1},\dots,p_{n}) & = & \int_{\mathcal{M}^{n}}d\vec{y}\,\biggl\{\mathcal{O}[\varphi_{n}(y_{1},
\dots,y_{n})]-\sum_{i,\epsilon_{i}}\Theta(\epsilon_{i}y_{i})\mathcal{O}[\varphi_{n}(y_{1},\dots,
\epsilon_{i}\infty,\dots,y_{n})]_{c}\nonumber \\
 &  & -\sum_{i,j,\epsilon_{i},\epsilon_{j}}\Theta(\epsilon_{i}y_{i},\epsilon_{j}y_{j})
 \mathcal{O}[\varphi_{n}(y_{1},\dots,\epsilon_{i}\infty,\dots,\epsilon_{j}\infty,\dots,y_{n})]_{c}+\dots\\
 &  & -\sum_{\{i_{k},\epsilon_{k}\}}\Theta(\epsilon_{i_{1}}y_{i_{1}},\dots,\epsilon_{i_{k}}y_{i_{k}})
 \mathcal{O}[\varphi_{n}(y_{1},\dots,\epsilon_{i_{1}}\infty,\dots,\epsilon_{i_{k}}\infty,\dots,y_{n})]_{c}+\dots\biggr\}\nonumber 
\end{eqnarray}
where we denoted the integration for the moduli space as $\int_{\mathcal{M}^{n}}d\vec{y}=\prod_{i}\int_{-\infty}^{\infty}dy_{i}$
.

\section{Diagonal form factors and expectation values in finite volume}

In this section we generalize the previous analysis for finite volume.
We assume that the volume $L$ is asymptotically large and neglect
all exponentially small vacuum polarization effects. We start by recalling
the available results for the quantum theory and then develop the
classical finite volume form factors in parallel with Section 2.

\subsection{Finite volume diagonal form factors}

We analyze a quantum field theory in a large volume $L$ and focus
on the leading (polynomial) finite size correction of the expectation
values. In this approximation the finite and infinite volume form
factors differ only by the normalization of states 
\cite{Pozsgay:2007kn}. The finite volume
states $\vert p_{1},\dots,p_{n}\rangle_{L}$ are eigenstates of energy
and momentum with the eigenvalues
\begin{equation}
P\vert p_{1},\dots,p_{n}\rangle_{L}=\sum_{k=1}^{n}p_{k}\vert p_{1},\dots,p_{n}\rangle_{L}\quad;
\qquad H\vert p_{1},\dots,p_{n}\rangle_{L}=\sum_{k=1}^{n}E(p_{k})\vert p_{1},\dots,p_{n}\rangle_{L}
\end{equation}
which are formally the same as the ones in infinite volume. The basic
difference is that a finite volume state is symmetric in the momenta,
and the momenta are quantized in a volume-dependent way by the Bethe-Yang
equation
\begin{equation}
e^{ip_{k}L}\prod_{j:j\neq k}S(p_{k},p_{j})=1\quad;\quad k=1,\dots,N
\end{equation}
 In practice, we take the logarithm of this equation
\begin{equation}
\Phi_{k}=p_{k}L-i\sum_{j:j\neq k}\log S(p_{k},p_{j})=2\pi I_{k}\label{eq:LogBY}
\end{equation}
and use the quantization numbers $\{I_{k}\}$ to label finite volume
states $\vert p_{1},\dots,p_{n}\rangle_{L}\equiv\vert I_{1},\dots,I_{n}\rangle$.
Due to the discreteness of the finite volume spectrum the states are
normalized to Kronecker $\delta$-functions: 
\begin{equation}
\langle J_{m},\dots,J_{1}\vert I_{1},\dots,I_{n}\rangle=\delta_{n,m}\delta_{I_{1}J_{1}}\dots
\delta_{I_{n}J_{n}}
\end{equation}
in contrast to the infinite volume states which are normalized to
Dirac $\delta$ functions. Both the finite and infinite volume states
form complete bases and we can relate them for large volumes by comparing
the resolution of the identity. For large volumes the momentum eigenstates
are very dense and we can change variables $\{p_{i}\}\to\{I_{i}\}$
via eq. (\ref{eq:LogBY}) leading to the relation 
\begin{equation}
\vert p_{1},\dots,p_{n}\rangle_{L}=\mathcal{N}\vert p_{1},\dots,p_{n}\rangle\quad;\qquad\mathcal{N}^{-1}
=\sqrt{\prod_{i<j}S(p_{i},p_{j})\rho_{n}(p_{1},\dots,p_{n})}\label{eq:FVnorm}
\end{equation}
Here the density of states is defined by the Jacobian: 
\begin{equation}
\rho_{n}(p_{1},\dots,p_{n})=\det\left[\Phi_{ij}\right]\quad;\qquad\Phi_{ij}=E(p_{i})\frac{\partial\Phi_{j}}
{\partial p_{i}}=\bigl(E(p_{i})L+\sum_{k=1}^{n}\phi_{ik}\bigr)\delta_{ij}-\phi_{ij}
\end{equation}
We also included the multiparticle S-matrix to compensate the order
dependence of the infinite volume state. We denoted the derivative
of the phase of the S-matrix with respect to the first argument as
\begin{equation}
\phi_{jk}=\phi(p_{j},p_{k})=-iE(p_{j})\frac{\partial}{\partial p_{j}}\log S(p_{j},p_{k})
\end{equation}
The derivative wrt. to the second argument is related to $\phi_{jk}$
by unitarity: $-iE(p_{k})\frac{\partial}{\partial p_{k}}\log S(p_{j},p_{k})=-\phi_{kj}$. 

Using the finite volume norm of states Saleur suggested an expression
for the finite volume expectation value in terms of the infinite volume
connected diagonal form factors \cite{Saleur:1999hq}\footnote{Similar formula was proposed 
for symmetric diagonal form factors in \cite{Pozsgay:2007gx} and 
proved later in \cite{Pozsgay:2013jua}.}:

\begin{eqnarray*}
\,_{L}\langle p_{n},...,p_{1}\vert\mathcal{O}\vert p_{1},\dots p_{n}\rangle_{L} 
& = & \frac{1}{\rho\{1,...,n\}}\sum_{A}\bar{\rho}\{A\}F_{\vert\bar{A}\vert}\{\bar{A}\}
\end{eqnarray*}
\begin{equation}
=\frac{F_{n}+\sum_{i}\bar{\rho}\{i\}F_{n-1}\{1,..,\hat{i},..n\}+\sum_{i,j}\bar{\rho}\{i,j\}
F_{n-2}\{1,..,\hat{i},..,\hat{j},..,n\}+\dots}{\rho\{1,..,n\}}
\label{eq:Fcdiag}
\end{equation}
where 
\begin{equation}
\bar{\rho}\{i_{1},\dots,i_{m}\}=\mbox{det}_{jk}\left[\Phi_{i_{j}i_{k}}\right]
\end{equation}
is the determinant of the minor of the Jacobi matrix involving the
set of labels $\{i_{1},\dots,i_{m}\}$. In particular, for one and
two particles we have

\begin{equation}
_{L}\langle p\vert\mathcal{O}\vert p\rangle_{L}=\frac{1}{\rho_{1}(p)}(F_{1}(p)
+\rho_{1}(p)F_{0})\quad;\qquad\rho_{1}(p)=EL\label{eq:rho1}
\end{equation}

\begin{equation}
_{L}\langle p_{2},p_{1}\vert\mathcal{O}\vert p_{1},p_{2}\rangle_{L}
=\frac{F_{2}(p_{1},p_{2})+\bar{\rho}_{1}(p_{1})F_{1}(p_{2})+\bar{\rho}_{1}(p_{2})F_{1}(p_{1})
+\rho_{2}(p_{1},p_{2})F_{0}}{\rho_{2}(p_{1},p_{2})}
\end{equation}
where 
\begin{equation}
\rho_{2}(p_{1},p_{2})=L^{2}E_{1}E_{2}+L(\phi_{12}E_{2}+\phi_{21}E_{1})\:
;\quad\bar{\rho}_{1}(p_{1})=E_{1}L+\phi_{12}\:;\quad\bar{\rho}_{1}(p_{2})=
E_{2}L+\phi_{21}\label{eq:rho2}
\end{equation}
and $E_{i}=E(p_{i})$. 

The expression (\ref{eq:Fcdiag}) for the finite volume expectation
values are valid upto exponentially small corrections. It incorporates
all polynomial correction in $L^{-1}$, which come from two sources.
Its explicit dependence sits in the norm of the states, while implicitly
it depends on $L$ via the momenta, which satisfy the Bethe-Yang equation
(\ref{eq:LogBY}). Observe that this expression is finite and provides
a regularization of the analogous infinite volume formula (\ref{diagvsff}).

\subsection{Classical limit of expectation values}

Recall that the expectation value can be thought of as the quantum
average of the operator $\mathcal{O}(\hat{\varphi}(x,t))$ in a finite
volume energy-momentum eigenstate. The classical analogue of this
formula should be in which we integrate the function $\mathcal{O}(\varphi(x,t))$
over the moduli space of the classical finite volume solutions with
the same energy and momentum. Similarly how the finiteness of the
volume regularized the quantum average, the classical integral is
finite, too. The quantum formula (\ref{eq:Fcdiag}) expresses this
finite average in terms of the infinite volume diagonal form factors
and the sub-densities $\bar{\rho}_{k}$. In an analogous way we express
the classical average in terms of the classical diagonal form factors
and the classical limit of the sub-densities $\bar{\rho}_{k}$. We
start by constructing the finite volume multiparticle solutions and
by determining their moduli space. We then rewrite the classical average
in terms of the classical diagonal form factors.

\subsubsection{Classical solutions and their moduli space}

The main difference between the infinite and finite volume solutions
is that the moduli space of the latter is finite. Let us analyze it
with increasing particle numbers.

\paragraph*{Vacuum}

The vacuum solution is automatically periodic and doesn't have any
moduli parameter.

\paragraph*{1-particle }

\begin{figure}
\begin{centering}
\includegraphics[width=4cm]{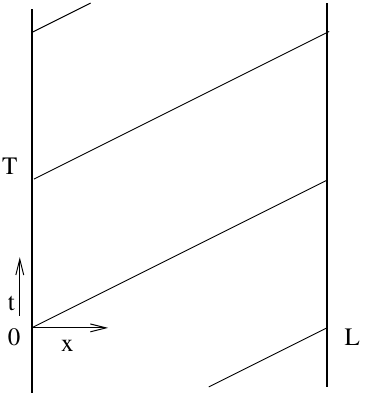}~~~~\includegraphics[width=4cm]{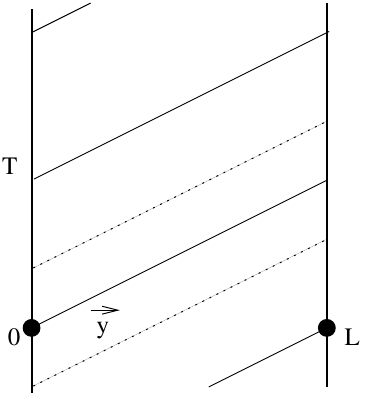}
\par\end{centering}

\protect\caption{The 1-particle trajectory in finite volume is $x(t)=v_{1}t+x_{1}-nL$,
where $x$ being understood modulo $L$. The moduli parameter shifts
the solution both in space and in time and its periodicity is $Y_{1}=E_{1}L$. \label{fig:1pfv} }
\end{figure}

The finite volume one particle solution is usually very complicated
and incorporates exponentially small finite size corrections. As we
focus only on the polynomial correction in $L^{-1}$ the exact solution
can be approximated by the infinite volume solution. In this approximation
the particles can be considered pointlike and we merely continue the
particle's trajectory periodically as shown on Figure \ref{fig:1pfv}.

The periodicity of this solution in time is $T_{1}=\frac{L}{v_{1}}$
and within each time period the finite volume 1-particle solution
is 
\begin{equation}
\varphi_{1}(x,t,y_{1})_{L}=\varphi_{st}(E_{1}x-y_{1}-p_{1}(t-nT_{1}))
\end{equation}
with some appropriately chosen $n$. The time/space periodicity translates
into the $y$- periodicity on the moduli space as:
\begin{equation}
y_{1}\equiv y_{1}+Y_{1}\quad;\qquad Y_{1}=p_{1}T_{1}=E_{1}L\equiv\rho_{1}^{c}
\end{equation}
Denoting the shift vector $y_{1}\to y_{1}+Y_{1}$, by $\Delta_{1}y=Y_{1}$,
the finite volume moduli space is the factor space 
\begin{equation}
\mathcal{M}_{L}^{1}=\frac{\mathcal{M}^{1}}{\Delta_{1}y}
\end{equation}
which can be chosen to be the interval $[0,Y_{1}]$. Clearly this
moduli space is finite.

\paragraph*{2-particle}

\begin{figure}
\begin{centering}
\includegraphics[width=8cm]{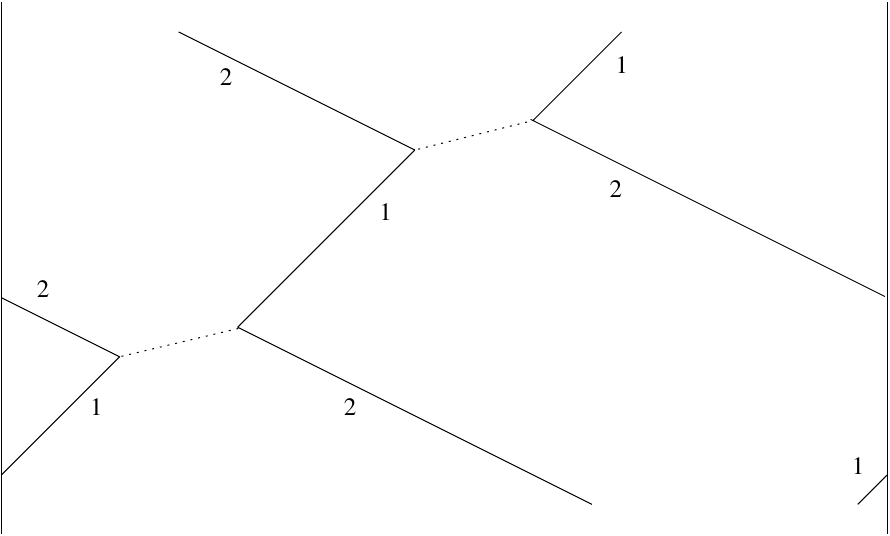}
\par\end{centering}

\protect\caption{Approximate finite volume $2$-particle solution: near the interaction
pont we use the infinite volume 2-particle-, while away of them, the
periodically continued infinite volume 1-particle solutions. \label{fv2psol}}
\end{figure}

The exact finite volume two particle solution is usually very complicated,
but we can easily construct a good approximate solution from the infinite
volume two particle solution as follows: we take two free particles
which travel as $x_{i}=v_{i}t+x_{i}^{-}$ and are well separated.
(In a large volume it is always possible). This can be approximated
by two 1-particle solutions. When the particles get close to each
other we replace this solution with the infinite volume 2-particle
solution. After the collision process, modeled by the two particle
solution, the particles are far away form each other and the 1-particle
approximation is correct again. However, the two trajectories are
now shifted as $x_{1}=v_{1}t+x_{1}^{-}+\Delta_{12}x$ and $x_{2}=v_{2}t+x_{2}^{-}+\Delta_{21}x$.
When any of these ``outgoing'' particles reaches the periodicity
border, $0$ or $L$, it will come back from the other side and together
with the other particle form a separated two particle initial state
similar we started with. We then repeat the previous scattering process
and by following this procedure we build up an approximate finite
volume 2-particle solution: Near the interaction pont we use the infinite
volume 2-particle-, while away from them, the infinite volume 1-particle
solutions as show on Figure \ref{fv2psol}. We denote this solution
as $\varphi_{2}(x,t;y_{1},y_{2})_{L}$ where $y_{1}$ and $y_{2}$
are related to the original coordinates $(x_{1}^{-},x_{2}^{-})$ of
the particles. 

Next we should understand the structure of the finite volume moduli
space. To parametrize this space we use the $y_{1}$ and $y_{2}$
shifts of the individual particles' locations as we used in the infinite
volume case. We search for such transformations on $y_{1},y_{2}$
which leave the finite volume solution invariant. We are going to
factor out with these transformations. In the 1-particle case we simply
moved the particle around the volume, which lead to the periodicity.
This is similar how we move a particle in the BY equation at the quantum
level. Let us mimic this transformation for the two particle case.
If we shift particle 1 to the right then it passes through particle
2 and comes back on the other side. Due to the interaction the periodicity
for particle $1$ is shorthand by the space-displacement as $L_{1}=L-\Delta_{12}x$.
The analogue periodicity in the moduli space is 
\begin{equation}
Y_{1}=E_{1}L-E_{1}\Delta_{12}x=E_{1}L+\phi_{12}^{c}
\end{equation}
This shift, however, does not leave the two particle configuration
invariant, because it is not the periodicity of the classical solution
(trajectory). The reason is that having passed through particle 2
it suffered a $\Delta_{21}x$ displacement thus for the full periodicity
we have to move back particle 2 by $-\Delta_{21}x$. Consequently,
the full periodicity is the simultaneous shifts on the plane 
\begin{equation}
(y_{1},y_{2})\to(y_{1}+Y_{1},y_{2}-\phi_{21}^{c})=(y_{1},y_{2})+(\Delta_{1}y_{1},
\Delta_{1}y_{2})\quad;\qquad\Delta_{1}y=(E_{1}L+\phi_{12}^{c},-\phi_{12}^{c})
\end{equation}
Similarly we can move also particle 2 into the right direction around
the circle. First we leave on the right and appear on the left and
then pass through particle 1. As $\Delta_{21}x$ is negative the effective
periodicity is shorthand to be $L-\vert\Delta_{21}x\vert$ or in the
moduli space to 
\begin{equation}
Y_{2}=E_{2}L+E_{2}\Delta_{21}x=E_{2}L+\phi_{12}^{c}
\end{equation}
Now passing particle 2 from the left through particle 1 the displacement
of particle $1$ is $-\Delta_{12}x$ which we compensate by adding
$\Delta_{12}x$. The full periodicity shift in the moduli space is
then 
\begin{equation}
(y_{1},y_{2})\to(y_{1}-\phi_{12}^{c},y_{2}+Y_{2})=(y_{1},y_{2})+(\Delta_{2}y_{1},
\Delta_{2}y_{2})\quad;\qquad\Delta_{2}y=(-\phi_{12}^{c},E_{2}L+\phi_{12}^{c})
\end{equation}
The finite volume moduli space is obtained by factoring out the infinite
volume moduli space by the two shift transformations 
\begin{equation}
\mathcal{M}_{L}^{2}=\frac{\mathcal{M}^{2}}{\{\Delta_{1}y,\Delta_{2}y\}}
\end{equation}
 The volume of this moduli space is finite, 
\begin{equation}
Vol_{2}=\rho_{2}^{c}=\mbox{det}[\Delta_{1}y,\Delta_{2}y]=L^{2}E_{1}E_{2}
+L(\phi_{12}^{c}E_{2}+\phi_{12}^{c}E_{1})
\end{equation}
and is nothing but the classical limit, $\rho^{c}$, of the density
of states (\ref{eq:rho2}).

\paragraph*{n-particle}

The finite volume approximate $n$-particle solution is constructed
as follows: we start with $n$ separated straight lines at $t=0$
with trajectories $x_{i}=v_{i}t+x_{-}^{i}$. The corresponding $n$-particle
solution is approximated by the sum of the one-particle solutions.
Whenever $k$ particles' lines approach each other (within the interaction
distance $\Delta_{ij}x$) we replace the sum of the $k$ one particle
solution with the infinite volume $k$-particle solution. We do this
construction on the cylinder (i.e. in a periodic way). We denote this
approximate finite volume solution by $\varphi_{n}(x,t;y_{1},\dots,y_{n})_{L}$. 

In order to determine the moduli space we analyze the symmetry of
the configuration. Let us move the $i^{th}$ particle around the cylinder.
When we pass particle $j$ we use the two particle scattering, so
the $i^{th}$ particle suffers a $\Delta_{ij}x$, while the $j^{th}$
particle a $\Delta_{ji}x$ displacement. In the moduli parameter we
multiply $x_{i}$ by $E_{i}$: $y_{i}=E_{i}x_{i}$. The simultaneous
transformation (shifts) in the moduli space which leaves the configuration
invariant is
\begin{equation}
i^{th}:\quad(y_{1},\dots,y_{i},\dots,y_{n})\to(y_{1}+\Delta_{i}y_{1},,
\dots y_{i}+\Delta_{i}y_{i},\dots,y_{n}+\Delta_{i}y_{n})
\end{equation}
\begin{equation}
\Delta_{i}y=(-\phi_{i1}^{c},\dots,Y_{i},\dots,-\phi_{in}^{c})\quad;
\qquad Y_{i}=LE_{i}+\sum_{j:j\neq i}\phi_{ij}^{c}
\end{equation}
The finite volume moduli space is the infinite volume moduli space
factored out by all the $n$ shift vectors
\begin{equation}
\mathcal{M}_{L}^{n}=\frac{\mathcal{M}^{n}}{\{\Delta_{1}y,\dots,\Delta_{n}y\}}
\end{equation}
 The volume of the phase space is the classical limit of $\rho_{n}$:
\begin{equation}
Vol_{n}=\rho_{n}^{c}=\mbox{det}[\Delta_{1}y,\dots,\Delta_{n}y]
\end{equation}

\subsubsection{Classical averages}

Similarly to the infinite volume case the quantum average of the operator
$\mathcal{O}(\hat{\varphi}(x,t))$ corresponds in the classical limit
to the average of the function $\mathcal{O}(\varphi(x,t))$ over the
finite volume moduli space of classical solutions. We express these
finite quantities in terms of the infinite volume form factors and
finite subvolumes of the moduli space.

\paragraph*{Vacuum}

As the vacuum solution is the same in finite and infinite volumes
the corresponding form factor is also the same $\mathcal{O}(\varphi_{0})$.
To simplify formulas we assume in the following that the observable,
$\mathcal{O}$, does not have any vacuum value.

\paragraph*{1-particle}

\begin{figure}
\begin{centering}
\includegraphics[width=5cm]{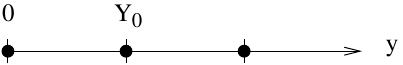}
\par\end{centering}

\protect\caption{Finite volume moduli space of $1$-particle solutions. It is periodic
with period $Y_{0}$. Black dots indicate the region where the $1$-particle
average (\ref{eq:1pav}) collects its contribution. \label{1pfvmod} }
\end{figure}

The classical expectation value of the function $\mathcal{O}(\varphi)$
in a 1-particle state with momentum $p$ is its average over the moduli
space of the finite volume solution 
\begin{equation}
_{\, L}\langle p\vert\mathcal{O}\vert p\rangle_{L}^{c}=\frac{1}{Y_{1}}
\int_{0}^{Y_{1}}dy_{1}\,\mathcal{O}(\varphi_{1}(x,t,y_{1})_{L})\label{eq:1pav}
\end{equation}
The main difference compared to the infinite volume expression is
that it is finite by itself. In the following we express this quantity
in terms of the infinite volume classical form factor. Clearly the
expectation value is independent of the space-time coordinates $(x,t)$
thus we insert the operator at the origin $(0,0)$, where the 1-particle
solution is passing by. For operators without vacuum expectation value
the integral collects contributions around the origin (denoted by
a black circle on Figure \ref{1pfvmod}). 

The finite volume expectation value in terms of the infinite volume
form factor can be written as 
\begin{equation}
_{\, L}\langle p\vert\mathcal{O}\vert p\rangle_{L}^{c}=
\frac{1}{Y_{1}}\int_{-\frac{Y_{1}}{2}}^{\frac{Y_{1}}{2}}dy_{1}\,\mathcal{O}[\varphi_{1}(y_{1})_{L}]
=\frac{1}{Y_{1}}\int_{-\infty}^{\infty}dy_{1}\,\mathcal{O}[\varphi_{1}(y_{1})]=
\frac{F_{1}^{c}}{\rho_{1}^{c}}\label{eq:1pfvav}
\end{equation}
where we used the fact that the contribution comes from a local region
around the origin and extended the domain of integration to infinity.
We also used that in our approximation the infinite and the finite
volume solutions are the same. The difference between the two expressions
in (\ref{eq:1pfvav}) is exponentially small and can be neglected.
This finite volume classical average is exactly the classical limit
of the quantum finite volume expectation value (\ref{eq:rho1}).

\paragraph*{2-particle}

\begin{figure}
\begin{centering}
\includegraphics[width=8cm]{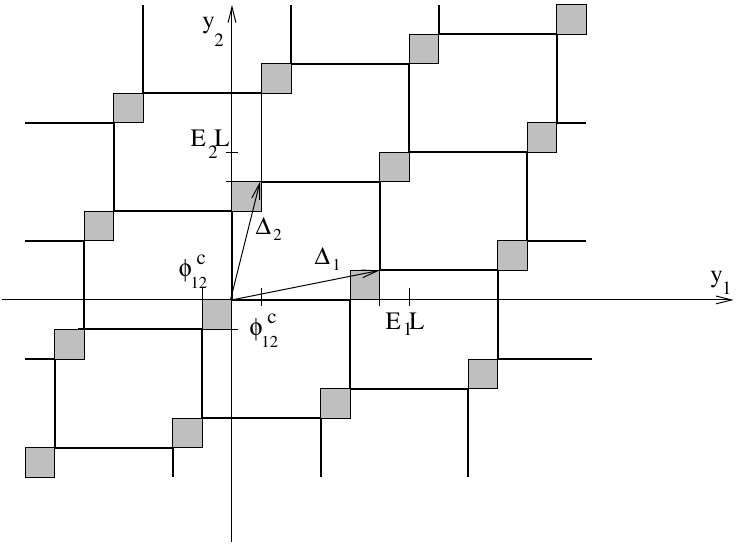}
\par\end{centering}

\protect\caption{Moduli space indicating the domains where the integral (\ref{eq:2pfvav})
collects its contributions. The picture is periodic with the shifts
$\{\Delta_{1}y,\Delta_{2}y\}$ to be factored out. \label{2ptfvcont}}
\end{figure}

The classical 2-particle expectation value is defined by averaging
the observable over the moduli space 
\begin{equation}
\,_{L}\langle p_{2},p_{1}\vert\mathcal{O}\vert p_{1},p_{2}\rangle_{L}^{c}=
\frac{1}{Vol_{2}}\int_{\mathcal{M}_{L}^{2}}dy_{1}dy_{2}\mathcal{O}
(\varphi_{2}(y_{1},y_{2})_{L})\label{eq:2pfvav}
\end{equation}
 The integral collects completely well-defined finite contributions
from the domain indicated on Figure \ref{2ptfvcont}.

\begin{figure}
\begin{centering}
\includegraphics[width=8cm]{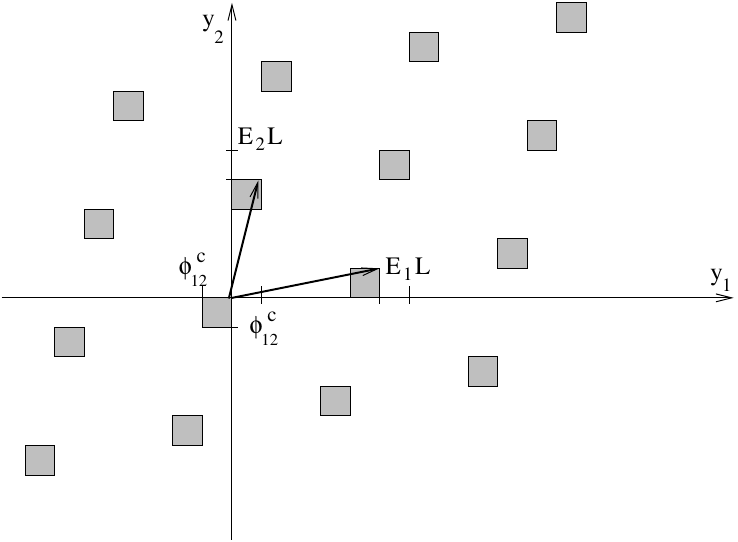}~~~~~\includegraphics[width=8cm]{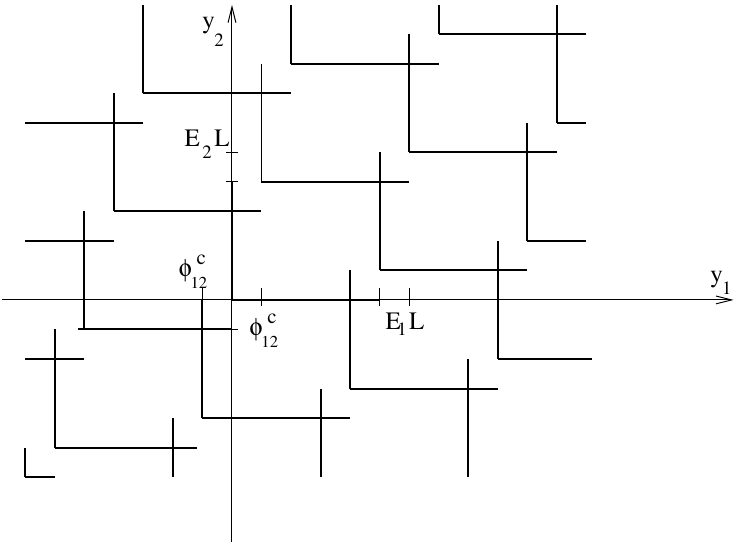}
\par\end{centering}

\protect\caption{$2$-particle and $1$-particle contributions of (\ref{eq:2pfvav})
in the moduli space. We indicated the shift vectors $\{\Delta_{1}y,\Delta_{2}y\}$
on the left figure explicitly. \label{2p1pfvcont}}
\end{figure}

\noindent This figure is the finite volume analogue of Figure \ref{2pffmodsub}. Similarly
to the infinite volume case let us separate the $2$-particle and
the $1$-particle contributions. It is indicated on Figure \ref{2p1pfvcont}.

\noindent In order to express the average in terms of the form factor
we subtract from $\mathcal{O}(\varphi_{2}(y_{1},y_{2})_{L})$ the
one particle contributions and add them back. We should be careful
with the subtraction as it has to be done in a way, which respects
the shift symmetries of the finite volume moduli space, $\{\Delta_{1}y,\Delta_{2}y\}$: 
\begin{equation}
\mathcal{O}(\varphi_{2}(y_{1},y_{2}))_{c}=\mathcal{O}(\varphi_{2}(y_{1},y_{2}))-
\Theta(y_{1})\Theta(Y_{1}-y_{1})\mathcal{O}(\varphi_{2}(\infty,y_{2}))-
\Theta(y_{2})\Theta(Y_{2}-y_{2})\mathcal{O}(\varphi_{2}(y_{1},\infty))
\end{equation}
 This formula is valid in the fundamental domain, and should be extended
periodically with the shifts $\{\Delta_{1}y,\Delta_{2}y\}$. Although
two subtracted pieces seem missing as compared to the infinite volume
expression, by shifting this function with the appropriate moduli
transformations the missing pieces can be recovered as 
\begin{equation}
-\Theta(-y_{1})\mathcal{O}[\varphi_{2}(-\infty,y_{2})]-\Theta(-y_{2})
\mathcal{O}[\varphi_{2}(y_{1},-\infty)]
\end{equation}
With these subtractions the classical finite volume expectation value
is
\begin{eqnarray}
\,_{L}\langle p_{2},p_{1}\vert\mathcal{O}\vert p_{1},p_{2}\rangle_{L}^{c} 
& = & \frac{1}{Vol_{2}}\int_{\mathcal{M}_{L}^{2}}dy_{1}dy_{2}\biggl\{\mathcal{O}
(\varphi_{2}(y_{1},y_{2}))_{c}+\Theta(y_{1})\Theta(Y_{1}-y_{1})
\mathcal{O}(\varphi_{2}(\infty,y_{2}))\nonumber \\
 &  & \,\,\,\,\,\,\,\,\,\,\,\,\,\,\,\,\,\,\,\,\,\,\,\,
 +\Theta(y_{2})\Theta(Y_{2}-y_{2})\mathcal{O}(\varphi_{2}(y_{1},\infty))\biggr\}
\end{eqnarray}
As both the 2-particle and the 1-particle integrands are localized
we can extend the integration domains appropriately to infinity. The
integrand in the subtracted/added back pieces factorize in $y_{1}$
and $y_{2}$. As the two particle solutions reduce to the $1$-particle
solutions when a particle shifted to infinity $\varphi_{2}(\infty,y_{2})=\varphi_{1}(y_{2})$
the integration for $y_{2}$ give the infinite volume diagonal form
factor $F_{1}^{c}(p_{2})$, while the integration for $y_{1}$ with
the $\Theta$ gives only the respective volume $Y_{1}$. Putting everything
together we obtain 
\begin{equation}
\,_{L}\langle p_{2},p_{1}\vert\mathcal{O}\vert p_{1},p_{2}\rangle_{L}^{c}=
\frac{1}{\rho_{2}^{c}}\left(F_{2}^{c}(p_{1},p_{2})+Y_{1}F_{1}^{c}(p_{2})+
Y_{2}F_{1}^{c}(p_{1})\right)
\end{equation}
which is exactly the classical limit of formula (\ref{eq:rho2}).

\paragraph*{n-particle }

The n-particle classical finite volume averages can be defined as
\begin{equation}
\,_{L}\langle p_{n},\dots,p_{1}\vert\mathcal{O}\vert p_{1},\dots,p_{2}\rangle_{L}^{c}=
\frac{1}{Vol_{n}}\int_{\mathcal{M}_{L}^{n}}dy_{1}\dots dy_{n}\mathcal{O}
(\varphi_{n}(y_{1},\dots,y_{n})_{L})
\end{equation}
This can be expressed in terms of the infinite volume connected integrands
as 
\begin{eqnarray}
\,_{L}\langle p_{n},\dots,p_{1}\vert\mathcal{O}\vert p_{1},\dots,p_{2}\rangle_{L}^{c} 
& = & \frac{1}{Vol_{n}}\int_{\mathcal{M}_{L}^{n}}d\vec{y}\,\biggl\{\mathcal{O}(\varphi_{1\dots n}(y_{1},\dots,y_{n})_{c})\\
 &  & +\sum_{i}\Theta\{y_{i}\}\mathcal{O}(\varphi_{n}(y_{1},\dots,\infty,\dots,y_{n}))_{c}\nonumber \\
 &  & +\sum_{i,j}\Theta\{y_{i},y_{j}\}\mathcal{O}(\varphi_{n}(y_{1},\dots,\infty,\dots,\infty,\dots,y_{n}))_{c}+\dots\nonumber \\
 &  & +\sum_{\{i_{k}\}}\Theta\{y_{i_{1}},\dots,y_{i_{k}}\}\mathcal{O}(\varphi_{n}(\{y_{i_{1}},
 \dots,y_{i_{k}}\}\to\infty))_{c}+\dots\biggr\}\nonumber 
\end{eqnarray}
The contributions of the lower order terms are such that after implementing
the various shifts the infinite volume subtractions are locally restored.
In particular, it implies that the various $\Theta$ terms are the
characteristic functions of the orthogonal projections of the moduli
space to the relevant set of variables. For one coordinate it is 
\begin{equation}\Theta\{y_{i}\}=\Theta(y_{i})\Theta(Y_{i}-y_{i})=
\begin{cases}1 & \mbox{if }y_{i}=\alpha_{i}\Delta_{i}y_{i}\quad\mbox{for some }\alpha_{i}\in[0,1]\\
    0 & \mbox{otherwise}\end{cases}
\end{equation}
For two coordinates it reads as
\begin{equation}
    \Theta\{y_{i},y_{j}\}=\begin{cases}1 & \mbox{if }y_{i}=\alpha_{i}\Delta_{i}y_{i}
    +\alpha_{j}\Delta_{j}y_{i}\mbox{ and }y_{j}=\alpha_{i}\Delta_{i}y_{j}+\alpha_{j}\Delta_{j}y_{j}
    \quad\mbox{for some }\alpha_{i},\alpha_{j}\in[0,1]\\
    0 & \mbox{otherwise}
\end{cases}
\end{equation}
while in general as
\begin{equation}\Theta\{y_{i_{1}},\dots,y_{i_{k}}\}=\begin{cases}1 & 
    \mbox{if for all} \; a=1,\dots, k: y_{i_{a}}=
    \sum_{j=1}^{k}\alpha_{j}\Delta_{i_{j}}y_{i_{a}}\quad\mbox{for some }
    \alpha_{i_{1}},\dots,\alpha_{i_{k}}\in[0,1]\\
    0 & \mbox{otherwise}
\end{cases}		  
\end{equation}
 By performing the
integral the integrand factorizes into the classical connected form
factors in one set of variables and the various classical densities
in the complementer set of variables leading to the formula, in which
the average of the observable $\mathcal{O}$ over the moduli space
of the classical $n$- particle solution can be written as 
\begin{equation}
\,_{L}\langle p_{n},\dots,p_{1}\vert\mathcal{O}\vert p_{1},\dots,p_{2}\rangle_{L}^{c}=
\frac{F_{n}^{c}(p_{1},\dots,p_{n})+\sum_{i}\rho_{1}^{c}(p_{i})F_{n-1}^{c}(p_{1},\dots,\hat{p}_{i},\dots,p_{n})+\dots}{\rho_{n}^{c}}
\end{equation}
which is the classical analogue of formula (\ref{eq:Fcdiag}).

\section{Some comments on HHL correlation functions}

As indicated in the introduction, it was the computation of Heavy-Heavy-Light correlation functions that was our main
motivation for developing the formalism of classical computation of finite volume expectation values
and diagonal form factors. In \cite{Bajnok:2014sza} we conjectured an identification between OPE coefficients
for `symmetric' operators i.e. when the two heavy operators are conjugate to each other, and 
diagonal form factors/finite volume expectation values.

Indeed, in the case where the heavy operator has charges only on the $S^5$,
the 2-point correlation function of the heavy state is
\eq
\label{e.shifted}
x_{\tau_0}(\tau)=R \tanh \kap (\tau-\tau_0) \quad\quad z_{\tau_0}(\tau) =\f{R}{\cosh \kap (\tau-\tau_0)} \quad\quad\text{and}
\quad X^I_{\{y^I\}}(\sg,\tau)
\eqx
where the solution on the $S^5$ also depends on its own set of moduli $\{y^I\}$ ($n$ moduli for an $n$-particle state).
We see that there is an additional moduli $\tau_0$ which is the relative time shift between the $AdS$ geodesic and the
solution on the $S^5$. The modified prescription for HHL correlators proposed in \cite{Bajnok:2014sza} is
\eq
\label{e.chhli}
C_{H\!H\!L}=const \cdot \lim_{T\to \infty} \f{1}{T} \int_{-T/2}^{T/2} d\tau_0 \int_{moduli/\Gamma}
\int d\tau d\sg V_L\left[ x_{\tau_0}(\tau), z_{\tau_0}(\tau),
X^I_{\{y^I\}}(\sg,\tau) \right]
\eqx
where we restricted ourselves to the case of conjugate heavy operators. Here we implicitly assume that the contribution
of the heavy vertex operators in the diagonal case will not have any moduli dependence and thus will not contribute 
to the above expression. Once we deal with the $\tau_0$ integral, which is usually trivial, the remaining integral
over the moduli space of finite volume classical solutions has exactly the same structure as the integral appearing
in the computation of finite volume classical expectation values discussed extensively in the previous section.
Thus one can adopt the decomposition into diagonal form factors obtained above also to this case\footnote{In Appendix~B we 
discuss a minor subtlety which is nevertheless harmless.}.
We conjectured that such general decomposition extends also beyond the classical case. This has recently been
verified at weak coupling and in the hexagon approach in \cite{Hollo:2015cda,Jiang:2015bvm,Jiang:2016dsr}.

In order to illustrate this formula, let us apply it to an interesting class of scalar operators including both
supergravity and massive short string excitations for which the vertex operators are known explicitly in
the classical limit. This family was introduced in \cite{Roiban:2009aa} and the vertex operators are 
\eq
\left(\f{x^2+z^2}{z}\right)^{-\Dl_L} \left[ \dw X^K \dwb X^K \right]^r
\eqx 
Since the AdS part factorizes, the $\tau_0$ integral can be easily carried out an one is left
with
\eq
C_{H\!H\!L} \propto \int_{moduli/\Gamma} \int d\tau d\sg  \left[ \dw X^K \dwb X^K \right]^r
\eqx
Now, specializing to the heavy solution to be contained in the $S^2 \subset S^5$, we can use Pohlmeyer reduction
formula to identify
\eq
\dw X^K \dwb X^K =\cos \beta \phi
\eqx
where $\phi$ is a sine-Gordon field. Thus one can reduce the computation of this class of HHL correlators to
diagonal form factors of the operators $e^{ik\beta \phi}$ in the sine-Gordon theory, for which we gave some explicit
expressions in the previous sections.
Note that the full expression for the finite volume expectation value
will be different as the Bethe Ansatz factors will be different from the ones in sine-Gordon theory.

\section{Conclusions}

In the present paper we proposed a scheme for performing computations in the classical limit
of two classes of observables in integrable field theories: diagonal form factors
in infinite volume and finite volume expectation values of local operators.
A key ingredient of the proposal is an integration over the moduli space of classical
multiparticle solutions which correspond to a single multiparticle quantum state.
The integration over the infinite volume moduli space is divergent which in fact mimics
the structure of divergences in infinite volume form factors in the diagonal limit.
The main contribution of this part of the paper is to provide a concrete prescription
for subtraction terms which lead to the infinite volume connected diagonal form factor which
is a perfectly finite quantity.

In the case of finite volume expectation values, although the moduli space has finite volume,
it has a nontrivial periodicity structure due to the time delays characteristic of
soliton scattering. We show that the relevant integral can be naturally evaluated in terms of
the classical diagonal form factors identified in the first part of the paper and
volume factors which turn out to be equivalent to subdeterminants of the Bethe ansatz equations.
In this way the known expression of finite volume expectation values in terms of diagonal form factors
are explicitly realized in terms of the proposed classical expressions. This is a very
nontrivial consistency check of the proposed expressions.

The relevance of the obtained results is twofold. On the one hand, the algorithm for the classical evaluation of
diagonal form factors may be an important and useful crosscheck of the full quantum expressions, as these are in fact
extremely complicated, as they would arise from a diagonal singular limit of a $2n$-particle form factor.
On the other hand, within the AdS/CFT correspondence the evaluation of Heavy-Heavy-Light OPE coefficients
reduces, as advocated in~\cite{Bajnok:2014sza}, to an integral over the moduli space of the
finite volume solution. This led to the conjecture spelled out in \cite{Bajnok:2014sza} that the HHL OPE
coefficients of `symmetric' operators are related to diagonal form factors through finite volume expectation values of the appropriate
part of the worldsheet vertex operator. The contribution of the present paper in this respect is to provide
a framework which works for any number of particles.

There are many interesting directions of future research. It 
would be particularly interesting to determine  the exact finite 
volume multiparticle solutions for an integrable QFT. Then one could
analyze the moduli space of these solutions and map its periodicity 
properties. A proper geometric quantization of this moduli space 
should lead to the Bethe-Yang equations. In the paper we provided 
explicit expressions for the classical limit of diagonal connected 
form factors with two particles for the exponential operators in the 
sine-Gordon theory. It 
would be challenging to evaluate the classical limit of the complicated quantum 
expression including multiple contour integrals to check our 
proposal. We calculated the explicit expressions for low powers of 
the exponential operators directly. It would be nice to find a 
closed expression for generic powers and to extend the results for 
higher multiparticle states. Work is in progress into these directions. 
The semiclassical finite volume form factors analyzed 
in \cite{Smirnov:1998kv} for the conformal case also revealed a 
connection with moduli space and Bethe-Ansatz equations. It would be 
very interesting to elaborate the connection between our results and 
\cite{Smirnov:1998kv} in order to find a unified description.

\bigskip

{\bf Acknowledgments.} RJ was supported by NCN grant 2012/06/A/ST2/00396 and ZB by a Lend\"ulet Grant and by
OTKA K116505. 
This work was supported by a Polish-Hungarian Academy of Science
cooperation.

\appendix

\section{Normalizations}

In this Appendix we comment on the normalizations of the states and
form factors. Clearly the normalization of multiparticle states affect
the form factor of all operators in a uniform way. 

In the paper we chose the normalization 
\begin{equation}
\langle p_{n},\dots,p_{1}\vert p'_{1},\dots,p'_{n}\rangle=\prod_{i=1}^{n}2\pi 
E(p_{i})\delta(p_{i}-p'_{i})\label{eq:relnorm}
\end{equation}
which is very natural from the relativistic point of view as it is
invariant under Lorentz transformations. It is nothing but $\delta$
normalization in rapidity space. In non-relativistic theories we could
alternatively normalize to $\delta$ functions in momentum space 
\begin{equation}
_{x}\langle p_{n},\dots,p_{1}\vert p'_{1},\dots,p'_{n}\rangle_{x}=\prod_{i=1}^{n}2\pi\delta(p_{i}-p'_{i})\label{eq:pnorm}
\end{equation}
which is indicated by a subscript $x$ as $2\pi\delta(p)=\int e^{ipx}dx$.
The diagonal matrix elements can be easily related in the two normalizations
\begin{equation}
\langle p_{1},...,p_{n}\vert\mathcal{O}\vert p_{n},\dots p_{1}\rangle=\prod_{i}E(p_{i})_{\, x}
\langle p_{1},...,p_{n}\vert\mathcal{O}\vert p_{n},\dots p_{1}\rangle_{x}
\end{equation}
 just as form factors 
\begin{equation}
F(p_{1},\dots,p_{n})=\prod_i E(p_{i})F^{x}(p_{1},\dots,p_{n})
\end{equation}
Changing the normalization of states will also change the density
of states to 
\begin{equation}
\rho_{n}^{x}(p_{1},\dots,p_{n})=\det\left[\Phi_{ij}^{x}\right]\quad;\qquad\Phi_{ij}^{x}=
\frac{\partial\Phi_{j}}{\partial p_{i}}=\bigl(L+\sum_{k=1}^{n}\phi_{ik}^{x}\bigr)\delta_{ij}-\phi_{ij}^{x}
\end{equation}
where
\begin{equation}
\phi_{jk}^{x}=\phi^{x}(p_{j},p_{k})=-i\frac{\partial}{\partial p_{j}}\log S(p_{j},p_{k})
\end{equation}
The finite volume expectation value is related to the Kronecker normalized
states and is thus normalization-independent:
\begin{eqnarray}
\,_{L}\langle p_{n},...,p_{1}\vert\mathcal{O}\vert p_{1},\dots p_{n}\rangle_{L} 
& = & \frac{1}{\rho^{x}\{1,...,n\}}\sum_{A}\bar{\rho}^{x}\{A\}F_{\vert\bar{A}\vert}^{x}
\{\bar{A}\}\label{eq:Fcdiag-1}
\end{eqnarray}
\begin{equation}
=\frac{F_{n}^{x}+\sum_{i}\bar{\rho}^{x}\{i\}F_{n-1}^{x}\{1,..,\hat{i},..n\}
+\sum_{i,j}\bar{\rho}^{x}\{i,j\}F_{n-2}^{x}\{1,..,\hat{i},..,\hat{j},..,n\}
+\dots}{\rho^{x}\{1,..,n\}}
\end{equation}
As in the classical limit 
\begin{equation}
\phi^{x}(p_{j},p_{k})\to\Delta_{jk}x
\end{equation}
it is more natural to think of the moduli space in terms of shifts of the
$x$-coordinate of the multiparticle solution. This coordinate is
dual to the momentum coordinate and it is easy to see from the normalization
change 
\begin{equation}
y_{i}=E_{i}x_{i}\quad;\qquad\int dy_{i}=E_{i}\int dx_{i}
\end{equation}
that the classical infinite volume form factor can be obtained as
\begin{eqnarray}
F_{n}^{x,c}(p_{1},\dots,p_{n}) & = & \prod_{i}\int_{-\infty}^{\infty}dx_{i}\,
\biggl\{\mathcal{O}[\varphi_{n}(x_{1},\dots,x_{n})]-\sum_{i,\epsilon_{i}}
\Theta(\epsilon_{i}x_{i})\mathcal{O}[\varphi_{n}(x_{1},\dots,\epsilon_{i}\infty,\dots,x_{n})]_{c}\nonumber \\
 &  & -\sum_{i,j,\epsilon_{i},\epsilon_{j}}\Theta(\epsilon_{i}x_{i})
 \Theta(\epsilon_{j}x_{j})\mathcal{O}[\varphi_{n}(x_{1},\dots,\epsilon_{i}
 \infty,\dots,\epsilon_{j}\infty,\dots,x_{n})]_{c}+\dots\nonumber \\
 &  & -\sum_{\{i_{k},\epsilon_{k}\}}\prod_{j=1}^{k}\Theta(\epsilon_{i_{j}}x_{i_{j}})
 \mathcal{O}[\varphi_{n}(x_{1},\dots,\epsilon_{i_{1}}\infty,\dots,
 \epsilon_{i_{k}}\infty,\dots,x_{n})]_{c}+\dots\biggr\}
\end{eqnarray}
What is nice about this normalization is that the classical analogue
of the Bethe-Yang equation has a direct geometric meaning. Indeed,
moving particle $i$ around the volume the $x$-space moduli parameters
change as 
\begin{equation}
i^{th}:\quad(x_{1},\dots,x_{i},\dots,x_{j},\dots,x_{n})\to(x_{1}+\Delta_{i}x_{1},,
\dots x_{i}+\Delta_{i}x_{i},\dots,x_{n}+\Delta_{i}x_{n})
\end{equation}
\begin{equation}
\Delta_{i}x=(-\Delta_{1i}x,\dots L-\sum_{j:j\neq i}\Delta_{ij}x,\dots,-\Delta_{ni}x)
\end{equation}
The volume of the coordinate-moduli space is indeed the classical
limit of the density of states 
\begin{equation}
\mathcal{M}_{L}^{n,x}=\frac{\mathcal{M}^{n}}{\{\Delta_{1}x,\dots,\Delta_{n}x\}}
\quad;\qquad Vol_{n}^{x}=\rho_{n}^{x,c}=\mbox{det}[\Delta_{1}x,\dots,\Delta_{n}x]
\end{equation}

\section{Connections to HHL 3-point functions}

In our previous paper we conjectured that the HHL three point functions
can be described by finite volume diagonal averages. In the strong
coupling (classical) limit we suggested a new way of calculating these
3-point functions by integrating the light vertex operator for the
moduli space of classical solutions. We explicitly checked and
connected these proposals by evaluating the two magnon matrix element
of the dilaton vertex operator. 

Our analysis for relativistic theories implies that the diagonal finite
volume matrix elements in the classical limit correspond to the integral
of the classical observable for the moduli space of classical solutions.
This, when applied to the HHL 3-point functions would imply the conjecture
for multiparticle state, however there is a caveat. Namely the AdS/CFT
correspondence is not described by a relativistic theory. Only its
classical limit can be mapped via the Pohlmeyer reduction to a relativistic
theory. In this map one also introduce a kind of gauge transformation,
which changes the effective size of the system and it is not quite
clear that the quantum-classical correspondence applies. In the following
we analyze the strong coupling (classical) limit of the quantum formulas
and show that it is consistent with the relativistic classical expressions.

We first recall that the strong coupling limit of the scattering matrix
is :
\begin{equation}
-i\log S(p_{1},p_{2})=-g(\cos\frac{p_{1}}{2}-\cos\frac{p_{2}}{2})
\log\frac{\sin^{2}(\frac{p_{1}-p_{2}}{4})}{\sin^{2}(\frac{p_{1}+p_{2}}{4})}
\end{equation}
where $g$ is the coupling constant, which goes to infinity. It is
related to the classical expression, which can be obtained by integrating
the time delay, by a gauge transformation and normalization \cite{Hofman:2006xt}:
\begin{equation}
-ig\log S^{c}(p_{1},p_{2})=-i\log S(p_{1},p_{2})-gp_{1}E_{2}
\end{equation}
In the finite volume formulas we need to calculate the density of
states, which is then expressed in terms of 
\begin{equation}
-i\frac{\partial\log S(p_{k},p_{j})}{\partial p_{k}}=g\Delta_{kj}x+gE_{j}
\end{equation}
via
\begin{equation}
\Phi_{ij}=\frac{\partial\Phi_{j}}{\partial p_{i}}=g(-\Delta_{1j}x-E_{j},
\dots,g^{-1}L+\sum_{k:k\neq j}(\Delta_{kj}x+E_{k}),\dots,-\Delta_{nj}x-E_{j})
\end{equation}
Introducing 
\begin{equation}
\tilde{L}=g^{-1}L+\sum_{i}E_{i}
\end{equation}
we can simply write 
\begin{equation}
g^{-1}\Phi_{ij}=(-\Delta_{1j}x,\dots,\tilde{L}+\sum_{j\neq i}\Delta_{ij}x,
\dots,-\Delta_{Ni}x)-E_{j}(1,\dots,1)
\end{equation}
The determinant of $\Phi_{ij}$ is the classical limit of the quantum
density, which we would like to relate to $\rho^{x,c}$ . The key
observation is that 
\begin{equation}
L^{-1}\mbox{det}[\Phi_{ij}]=g^{n}\tilde{L}^{-1}\rho^{x,c}
\end{equation}
This can be shown by simultaneous transformations on both matrices.
First, by subtracting the first column from each we get rid off the
extra $E_{j}$ terms everywhere except the first column, such that
the rest of the matrices coincides. In the second step we add each
row to the first. As a result, the first row will be zero except the
first element, which is $\tilde{L}-\sum_{i}E_{i}=g^{-1}L$ for $\mbox{det}[\Phi_{ij}]$,
while $\tilde{L}$ for $\rho^{x}$.

\providecommand{\href}[2]{#2}\begingroup\raggedright\endgroup

\end{document}